\RequirePackage{fix-cm}
\documentclass[11pt]{article}

\usepackage{fullpage}
\usepackage[small,bf]{caption}
\usepackage[section]{placeins}
\usepackage{array}
\usepackage{cite}
\usepackage{calc}
\usepackage{caption}
\usepackage{graphicx}
\usepackage{psfrag}
\usepackage{stfloats}
\usepackage{url}
\usepackage{booktabs,siunitx}
\usepackage{subfigure,relsize}
\usepackage{longtable}
\usepackage{amsmath,amsfonts,amssymb,amsbsy,bm,paralist,theorem,ifthen,color}
\usepackage{mathtools}
\usepackage{algorithmic,algorithm}
\usepackage[dvips]{epsfig}
\usepackage[dvips]{graphicx}
\usepackage{caption}
\usepackage[font={small}]{caption}

\definecolor{orange}{RGB}{255,107,0}

\definecolor{green}{RGB}{0,180,80}

\newcommand\Lc{\ensuremath{{\mathcal{L}}}}

\newcommand\yb{\ensuremath{{\bm y}}}

\newcommand\ub{\ensuremath{{\bm u}}}
\newcommand\Hb{\ensuremath{{\bm H}}}
\newcommand\hb{\ensuremath{{\bm h}}}
\newcommand\Ab{\ensuremath{{\bm A}}}
\newcommand\ab{\ensuremath{{\bm a}}}

\newcommand\Cb{\ensuremath{{\bm C}}}
\newcommand\cb{\ensuremath{{\bf c}}}

\newcommand\Db{\ensuremath{{\bm D}}}
\newcommand\eb{\ensuremath{{\bm e}}}

\newcommand\fb{\ensuremath{{\bm f}}}
\newcommand\Fb{\ensuremath{{\bm F}}}
\newcommand\Gb{\ensuremath{{\bm G}}}
\newcommand\gb{\ensuremath{{\bm g}}}
\newcommand\Ib{\ensuremath{{\bm I}}}

\newcommand\Sb{\ensuremath{{\bm S}}}

\newcommand\Ub{\ensuremath{{\bm U}}}
\newcommand\Rb{\ensuremath{{\bm R}}}
\newcommand\Mb{\ensuremath{{\bf M}}}

\newcommand\nb{\ensuremath{{\bf n}}}

\newcommand\Qb{\ensuremath{{\bm Q}}}

\newcommand\Vb{\ensuremath{{\bm V}}}
\newcommand\vb{\ensuremath{{\bm v}}}
\newcommand\Wb{\ensuremath{{\bm W}}}

\newcommand\zb{\ensuremath{{\bm z}}}
\newcommand\alphab{\ensuremath{{\bm \alpha}}}

\newcommand\mub{\ensuremath{{\bm \mu}}}

\newcommand\zerob{\ensuremath{{\bm 0}}}

\newcommand\Cbb{\ensuremath{{\mathbb{C}}}}

\newtheorem{Rmk}{Remark}

\allowdisplaybreaks[4]

\title{CSI Sensing from Heterogeneous User Feedbacks: A Constrained Phase Retrieval Approach}

\author{
	Lei~Li,~
	Xing Zeng,
	Ya-Feng Liu,
	Yanqing Xu,
	and Tsung-Hui~Chang
	\thanks{\smaller[1]T.-H. Chang (the corresponding author) is with the School of Science and Engineering, The Chinese University of Hong Kong, Shenzhen, China, and with the Shenzhen Research Institute of Big Data (email: tsunghui.chang@ieee.org). L. Li is with the School of Science and Engineering, The Chinese University of Hong Kong, Shenzhen, China, and with the Shenzhen Research Institute of Big Data (email: lei.ap@outlook.com). X. Zeng is with Huawei Technologies Ltd., Shanghai, China (e-mail: zengxing1@huawei.com). Y.-F. Liu is with the State Key Laboratory of Scientific and Engineering Computing, Institute of Computational Mathematics and Scientific/Engineering Computing, Academy of Mathematics and Systems Science, Chinese Academy of Sciences, Beijing 100190, China (e-mail: yafliu@lsec.cc.ac.cn). Y. Xu is with the School of Science and Engineering, The Chinese University of Hong Kong, Shenzhen, China, and also with the Shenzhen Research Institute of Big Data (e-mail: xuyanqing@cuhk.edu.cn).}
	\thanks{This work has been submitted to the IEEE for possible publication.  Copyright may be transferred without notice, after which this version may no longer be accessible.}
	}

\date{}

\begin{document}
	
	\maketitle
	\begin{abstract}		
		This paper investigates the downlink channel state information (CSI) sensing in 5G heterogeneous networks composed of user equipments (UEs) with different feedback capabilities. We aim to enhance the CSI accuracy of UEs only affording the low-resolution Type-I codebook. While existing works have demonstrated that the task can be accomplished by solving a phase retrieval (PR) formulation based on the feedback of precoding matrix indicator (PMI) and channel quality indicator (CQI), they need many feedback rounds. In this paper, we propose a novel CSI sensing scheme that can significantly reduce the feedback overhead. Our scheme involves a novel parameter dimension reduction design by exploiting the spatial consistency of wireless channels among nearby UEs, and a constrained PR (CPR) formulation that characterizes the feasible region of CSI by the PMI information. To address the computational challenge due to the non-convexity and the large number of constraints of CPR, we develop a two-stage algorithm that firstly identifies and removes inactive constraints, followed by a fast first-order algorithm. The study is further extended to multi-carrier systems. Extensive tests over DeepMIMO and QuaDriGa datasets showcase that our designs greatly outperform existing methods and achieve the high-resolution Type-II codebook performance with a few rounds of feedback.
		
		\vspace{0.5cm}
	\end{abstract}
	
	%

	%
	
	\section{Introduction} \label{sec:intro}
	
	With the coming age of massive multiple-input multiple-output (MIMO), the performance of wireless networks can be significantly boosted by multi-antenna techniques like beamforming and precoding. Most multi-antenna methods rely on accurate channel state information (CSI) at the base station (BS). 
	In frequency division duplex (FDD) systems,  due to the lack of reciprocity between uplink (UL) and downlink (DL) channels, the estimated DL CSI at the user equipment (UE) needs to be fed back to the BS.       
	However, both the DL CSI estimation and feedback are computation-consuming and resource-demanding.  
	Specifically,  with the increasing number of antennas and more stringent requirements of CSI accuracy, the overhead of CSI feedback proliferates. When more time-frequency resource is used for CSI feedback, the resource for the data transmission has to be decreased. As a result, the spectrum efficiency (SE) of data transmission is degraded, and it calls for highly efficient CSI sensing techniques.	
	
	The DL CSI estimated at a user equipment (UE) in practical cellular systems is usually quantized with a predefined codebook. Then a corresponding precoding matrix indicator (PMI) is fed back to the BS. The choice of codebook requires balancing the feedback overhead and the accuracy of CSI returned. Two kinds of codebooks are specified by the 3GPP \cite{TS_38214}-- the low-resolution Type-I codebook that has been widely applied in 4G systems and the Type-II codebook that is added to 5G to enhance the CSI accuracy. The Type-I codebook characterizes the CSI with a single beam and conveys the coarse information about the DL CSI. The Type-II codebook depicts the CSI with multiple discrete Fourier transform (DFT) rays and keeps the amplitude and phase information. Therefore, the CSI conveyed by the Type-II codebook is more accurate and enhances the data transmission performance effectively. System-level performance evaluation \cite{VTC19_sysSimu, VTC18_sysSimu} has demonstrated that up to 30\% performance improvement can be achieved by the Type-II codebook based beamforming over its counterpart on the Type-I codebook.  
	
	\begin{figure}[t] 
		\centering	
		\includegraphics[width=0.75\textwidth]{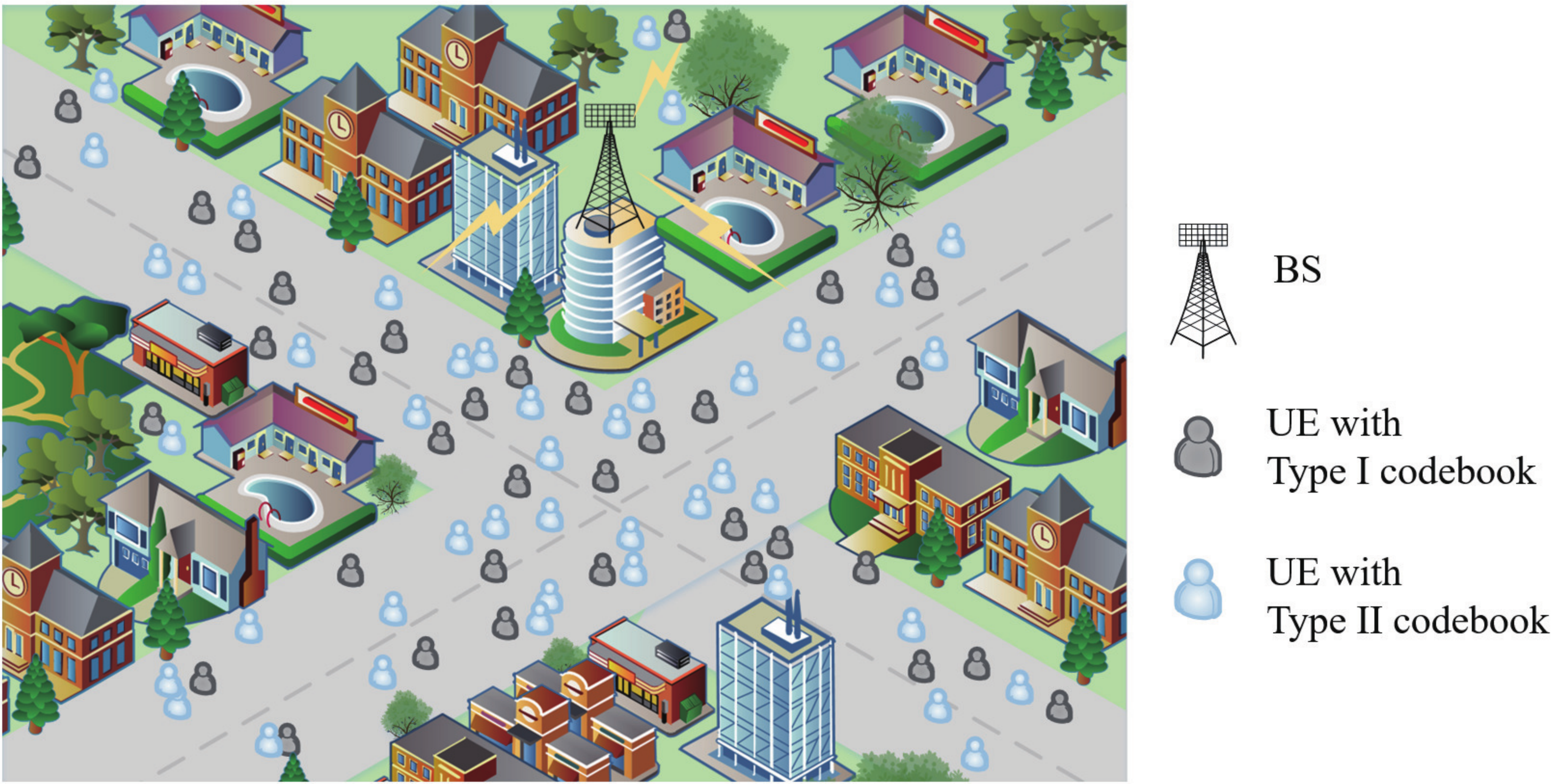}
		\caption{Illustration of a wireless network consisting of UEs with different CSI feedback capabilities. } 
		\label{fig:net}
	\end{figure}
	
	Though the Type-II codebook based on CSI feedback brings excellent performance enhancement, it has not entirely replaced the Type-I codebook. First, the accuracy of CSI improved with the Type-II codebook is at the price of a much higher overhead. Because of that, the Type-II codebook currently supports up to rank-2 transmission only, while up to rank-8 transmission is enabled by the Type-I codebook. On the other hand, the Type-I codebook has been extensively used in 4G systems for many years. With the fact that a considerable number of UEs from the 4G age are still widely in use, and most of them only support the Type-I codebook, it can be anticipated that they will coexist with more advanced UEs like 5G cellphones supporting the Type-II codebook for a long time, resulting in a heterogeneous network as illustrated in Fig. \ref{fig:net}.  
	
	In this paper, we consider such heterogeneous networks.  Our goal is to improve the accuracy of CSI reconstruction at the BS for these UEs that can afford the Type-I codebook only. In particular, we aim to utilize the PMI and channel quality indicator (CQI) of the Type-I codebook fed back from the target UE (TU) to recover its CSI with high accuracy.
	This CSI sensing task can be formulated as a phase retrieval (PR) problem via appropriate training signal design \cite{Sidiropoulos_twc15, LiKai_1, LiKai_2}.  
	However,  like the conventional PR, which requires many measurements for accurate recovery,  these methods require many
	feedbacks from the TU to recover the CSI well. 
	To effectively reduce the feedback overhead,  in this paper, we exploit the low-rank structure of the channel model and the PMI feedback to
	propose a novel dimension reduced and constrained PR (CPR) formulation, together with new precoding designs and efficient CSI recovery algorithms.
	\FloatBarrier
	\subsection{Related Works}
	
	There have been many efforts to reduce the feedback overhead in FDD systems in the literature. 
	Conventionally, compressive sensing (CS) based methods were widely used to reduce the CSI feedback overhead \cite{CS1, CS_est_mag10, CS_est_WCNC, CS2, CS_VLau}. The spatial correlation between closely-packed antennas was utilized, and an adaptive channel feedback protocol based on CS was proposed in \cite{CS1} to reduce the CSI feedback load, with 2D discrete cosine transform and Karhunen-Loeve transform used as bases. 
	The work \cite{CS2} considered the CSI estimation in FDD multi-user massive MIMO systems. A joint channel sparsity model was proposed, wherein the sparsity is divided into individual joint sparsity and distributed joint sparsity by different types of scattering. An algorithm based on orthogonal matching pursuit (OMP) was designed to reduce the training and feedback overhead for the CSI acquisition. Similar ideas were adopted in \cite{CS_VLau} and a sparse Bayesian learning framework was proposed for the joint channel estimation and user grouping. Although the designs in these works are shown to reduce the overhead effectively, they all have assumed that the UE can feedback the estimated CSI ideally without rate constraints and quantization.

	Recently, with the popularity of deep learning and its efficacy validated in areas like image processing and natural language processing, many research works explored deep learning (DL) methods to reduce the CSI feedback and recovery overhead. In \cite{CsiNet}, a CsiNet based on the auto-encoder was first proposed for the CSI compression and recovery. It improves the quality of CSI recovery effectively over the CS-based methods and without making any sparsity assumption. Inspired by this, the auto-encoder-likewise CSI feedback and recovery mechanisms were further investigated under more practical system settings like quantization, coding, and modulation to be considered \cite{A8, A9, A3, A6, A5, A12}. 
	Besides the consideration of the practical setting in the CSI feedback, other existing works tried to improve the compression efficiency via exploiting the temporal correlation of CSI \cite{A1,A2,A7}. Among these designs, the recurrent NN (RNN) and long-short-temporal-memory (LSTM) structures were adopted to exploit the temporal and frequency correlation of channels. 
	
	Although many schemes based on NN have been proposed to reduce the CSI feedback overhead, the feedback mechanism they rely on is not directly applicable to existing cellular systems. Moreover, most of these schemes only exploit temporal and frequency correlation. A few works have considered the spatial consistency, but are still limited on a small scale, i.e., the correlation of the CSI on different antennas of the same UE terminal \cite{A16,A19}.  
	
	It is worth pointing out that there exist several works \cite{Sidiropoulos_twc15, LiKai_1, LiKai_2} that share the same goal as our current work and target improving the CSI recovery at the BS with low-resolution feedback.  
	The work \cite{Sidiropoulos_twc15} designed a binary CSI feedback scheme, where the receiver returns a binary indicator to inform whether the received signal-to-noise ratio (SNR) is more significant than a preset threshold. A cutting plane-based algorithm was proposed to learn the channel covariance matrix from these feedbacks. While the CSI accuracy can be improved from multiple coarse feedbacks, the scheme in \cite{Sidiropoulos_twc15} cannot be applied to existing cellular systems with the PMI and CQI feedback mechanism. Comparatively, the designs in \cite{LiKai_1} and \cite{LiKai_2} also made use of the feedback PMI and CQI of the Type-I codebook. Still, the corresponding problem formulation and proposed schemes are different from our design. For instance, a volume minimization problem was formulated in \cite{LiKai_1} and a cutting plane method like that in \cite{Sidiropoulos_twc15} was proposed to search the channel covariance matrix. The work \cite{LiKai_2} formulated a 2-norm regularized PR problem for CSI sensing, and an alternating minimization algorithm was devised. However, both methods in \cite{LiKai_1, LiKai_2} still take a large number of feedbacks (e.g.,  more than $30$ rounds of feedbacks in \cite{LiKai_1}) to achieve comparable performance as the Type-II codebook. Besides, they neither exploit the spatial consistency nor consider the precoding design.

	\subsection{Contributions}
	
	In this paper,  like \cite{LiKai_1, LiKai_2},  we aim to design a DL CSI sensing scheme to enhance the CSI recovery accuracy of UEs that can afford the low-resolution Type-I codebook only. 
	The objective is to effectively reduce the feedback overhead of the TU while achieving a sensing performance as good as that of the high-resolution (Type-II) codebook in a computationally efficient manner.   
	
	To reduce the feedback overhead,  we propose exploring the wireless channel's low-rank structure,  which enables us to project the original high-dimensional CSI onto a low-dimensional space so that the BS only needs to recover a small number of unknown parameters from the TU feedback.  Since in the PR the required number of measurements is proportional to the number of parameters, this strategy can effectively decrease the number of TU feedbacks for accurate CSI sensing. 
	While such dimension reduction is possible,  how to construct the associated basis matrix at the BS is unknown. To address it, we exploit the spatial consistency of the CSI among geographically nearby UEs. Specifically,  as nearby UEs' channels experience similar reflectors and scatters, they share similar structures as the TU's channel.  It is noted that experiments have widely validated the spatial consistency of the CSI \cite{SC_meas1, SC_meas2, SC_meas3} and it is observed in many studies \cite{SC_obv1, SC_obv2}. 
	Recently, the spatial consistency feature was also emphasized in standards and channel models \cite{TR_38901, QuaDriGa}. 
	Based on the spatial consistency, we assume that the TU's CSI approximately lies in the same space as the CSI of nearby reference UEs (RUs). Therefore, as shown in Fig. \ref{fig:net},  it is possible to construct the basis for the TU's CSI from the Type-II PMI feedback of RUs.
	
	To achieve a promising CSI sensing performance,  we utilize the constructed low-rank CSI basis to design the training precoding matrix.  In particular, we propose a hybrid precoding structure that is a product of a Gaussian random matrix and the projection matrix for the low-rank CSI subspace.  The Gaussian random matrix is used for improving the sensing diversity in the PR \cite{PRIME_tsp16, LiKai_2}. In contrast, the projection matrix can align the training signal with the channel and thereby enhances the received SNR at the TU.  Compared to the current work in \cite{PRIME_tsp16, LiKai_2} which uses only Gaussian random precoders,  the proposed hybrid precoder can further boost the CSI sensing performance.
	Moreover,  we utilize the PMI fed back from the TU to characterize the feasible region that the CSI lies in,  which leads to a novel CPR formulation.  
	
	The formulated CPR problem is intricate to solve since it is not only non-convex in the objective function but also involves a large number of non-convex quadratic constraints.  
	For example, the problem could have thousands of constraints due to the Type-I codebook size (e.g., more than $1000$ codewords for the codebook with 32 CSI ports \cite{TS_38214}). 
	To address the challenges,  we propose a two-stage strategy.
	First,  we propose a simple algorithm to identify redundant constraints and construct a `minimal' effective constraint set (MECS) for the CPR problem.  
	Second,  we adopt the smoothed gradient descent ascent (SGDA) algorithm in \cite{ZJW} to solve the Lagrange dual problem of the CPR problem.  Our technical contributions are summarized as follows.
	
	\begin{enumerate}[1)]
		\item  Based on three novel ingredients, we propose in this paper a new CSI sensing scheme for achieving superior CSI recovery performance at the expense of considerably reduced feedback overhead for UEs that can afford the Type-I codebook only.  In particular, firstly, we exploit the spatial consistency of wireless channels to perform dimension reduction for the PR problem.  Secondly,  a hybrid precoder is proposed to enhance the CSI  sensing performance, and thirdly a constrained PR problem is formulated based on the PMI feedback.  By using the primal-dual (PD) optimization based algorithm in \cite{spawc22} (which we refer to as PD-EVD in the manuscript), we show that the proposed scheme significantly outperforms the existing works in \cite{LiKai_1, LiKai_2} and can recover accurate CSI with only a handful number of feedbacks.
		
		%
		
		\item  To solve the challenging CPR problem efficiently,  we propose the two-stage MECS-SGDA algorithm, which removes redundant constraints from the CPR problem first, followed by solving the simplified problem by the SGDA algorithm.  It will be shown that the MECS-SGDA algorithm can save at least a magnitude of time in solving the CPR problem compared to the PD-EVD algorithm.
		
		\item The proposed CSI sensing scheme is further extended to frequency selective fading channels and multi-carrier systems.  Rather than naively applying the previous CSI sensing scheme to each of the subcarriers,  we consider the channel correlation over different subcarriers and propose a new multi-carrier CSI sensing scheme that can provide even better performance.  
		
	\end{enumerate} 
	
	We remark that compared to the conference paper \cite{spawc22}, the current paper presents the new MECS-SGDA algorithm and the extension to multi-carrier systems. In addition, comprehensive simulation results based on the DeepMIMO and QuaDriGa datasets are presented to demonstrate the efficacy of the proposed methods.
	
	%
	%
	%
	
	{\bf Synopsis:} The system model and problem formulation are introduced in Section \ref{sec:model}. In Section \ref{sec:alg_design}, the  designs for dimension reduction, precoding and CPR modeling are elaborated in details. The algorithm design for the CPR problem is presented in Section \ref{sec:MECS_SGDA} and the extension to the multi-carrier systems is given in Section \ref{sec:mulC}.  Simulation results are given in Section \ref{sec:simu}. Finally, conclusions are drawn in Section {\ref{sec:conclusion}}.

	\section{System Model and Problem Formulation} \label{sec:model}

\begin{figure}[t] 
		\centering	
		\includegraphics[width=0.75\textwidth]{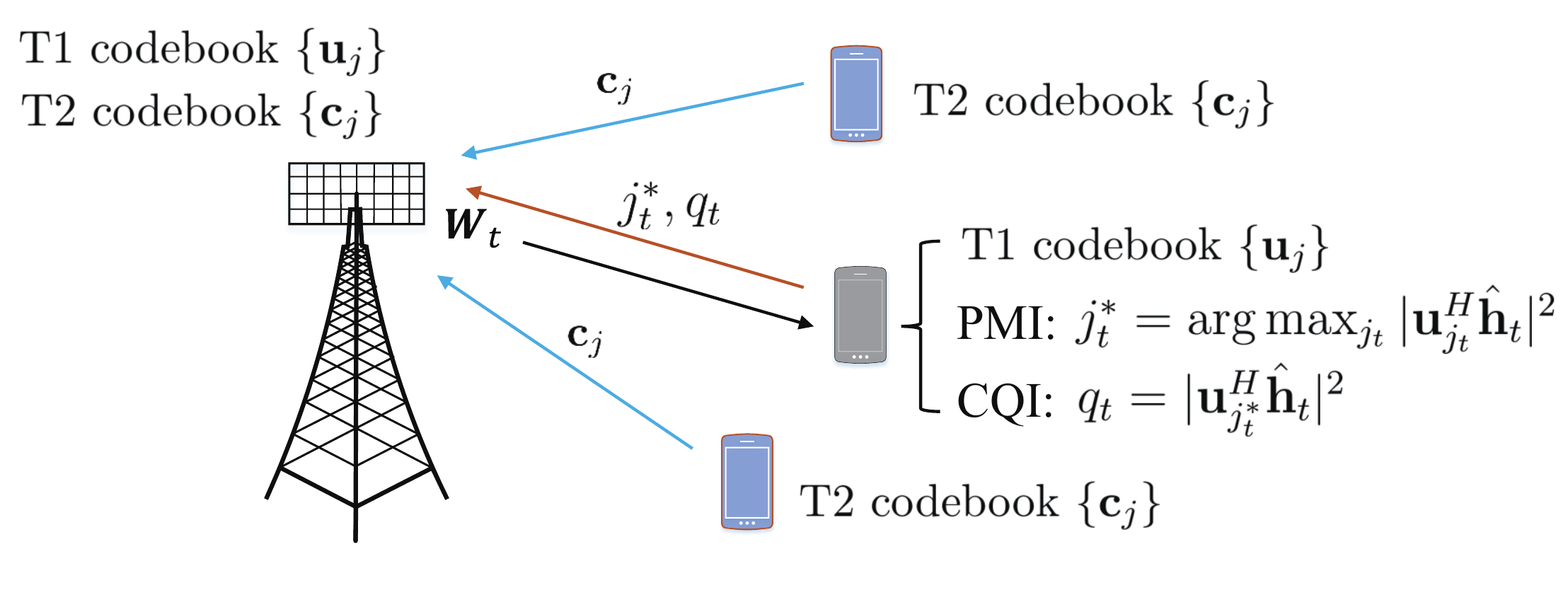}
		\caption{CSI sensing in a heterogeneous network, where the TU can only afford the low-resolution CSI feedback via the PMIs of Type-I codewords and CQIs, and nearby RUs return high-resolution CSI via the PMIs of Type-II codewords.} 
		\label{fig:system}
	\end{figure}

	As shown in Fig. \ref{fig:system}, we consider a heterogeneous wireless network with one BS communicating with multiple UEs. The BS is equipped with $M$ antennas. Each UE is equipped with a single antenna.  Here we assume flat fading channels,  and latter extend the work to the frequency-selective fading channels in Section \ref{sec:mulC}.  It is assumed that the TU can only return the coarse CSI via the PMI of Type-I codewords. At the same time, the nearby RUs are capable of producing the high-accuracy CSI via the PMI of Type-II codewords. In the work, by integrating the sensing data returned to the BS, the spatial consistency of the CSI among nearby users is exploited to improve the CSI sensing accuracy for the TU.

	The procedure of DL training and feedback between the BS and the TU is described as follows. Assume that $N_p \le M$ antenna ports are used for CSI sensing. Denote the DL pilot signal at the time slot $t$ as $\Sb_t \in \Cbb^{N_p \times N_p}, t= 1, \dots, T$. In the DL CSI estimation, the BS applies the precoding matrix $\Wb_t \in \Cbb^{M \times N_p}$. Assume that the (flat fading) channel $\hb \in \Cbb^{M}$ is quasi-static and is unchanged during the sensing period. The signal received  by the  TU  at time slot $t$ is expressed as
	\begin{equation} \label{eq:rx_model}
		\yb_t = \hb^H \Wb_t \Sb_t + \nb_t,
	\end{equation}
	where $\nb_t \sim \mathcal{CN}(\mathbf{0}, \sigma^2 \Ib_M)$ is the additive white Gaussian noise (AWGN). At the TU, without the knowledge of $\Wb_t$, only the effective CSI $\hat\hb_t$ can be estimated as 
	\begin{equation} \label{eq:CSI_eff}
		\hat\hb_t = \Wb_t^H \hb + \eb_t,
	\end{equation}
	where $\eb_t$ is the error of channel estimation. 
	Because returning the attained $\hat\hb_t$ to the BS directly is too resource consuming,  the TU in practice quantizes $\hat\hb_t$ through the Type-I codebook
	\begin{equation} \label{eq:T1_cw}
		\mathcal{U} \triangleq \{ \ub_j \in \Cbb^{N_p}, j = 1, \dots, n_{T1}\},
	\end{equation} 
	and selects a PMI that is the index of the best matched codeword as the feedback message.  Specifically, the PMI is determined by
	\begin{equation} \label{eq:T1_sel}
		j_t^* = \arg \max_{j = 1, \dots, n_{T1}} |\ub_j^H \hat\hb_t|.
	\end{equation}
	The CQI is also computed as 
	\begin{equation} \label{eq:CQI}
		q_t = |\ub_{j_t^*}^H \hat \hb_t|^2,
	\end{equation} 
	and both PMI and CQI are sent to the BS.
	
	When $N_p = M$ and $\Wb_t$ is an identity matrix $\Ib_M$, the codeword $\ub_{j_t^*}$ is simply an estimate of the DL CSI and can be directly used as the beamforming vector for data transmission. However, with a limited size, the Type-I codebook specified in \cite{TS_38214} only conveys the coarse information about the CSI $\hb$ and the performance is not satisfactory in general. 
	While the more sophisticated Type-II codeword can return the more accurate CSI, it needs much larger overhead. As depicted in the introduction section, not all the UEs can afford the Type-II codebook. Therefore, it raises an interesting question: \textit{Can a BS achieve a CSI estimate as good as the Type-II codeword using only the Type-I codeword feedbacks from the TU?} 
	
	In \cite{LiKai_2}, the authors showed that this is possible by solving the following PR problem
	\begin{equation} \label{eq:PR}
		\min_{\hb \in \Cbb^M} \sum_{t=1}^T \big(q_t - |\ub_{j_t^*}^H  \Wb_t^H\hb|^2 \big)^2.
	\end{equation}  
	According to theoretical results on the phase retrieval in \cite{PR_4N_1,PR_4N_2}, it typically requires $T\ge 4M$ measurements to retrieve the complex vector $\hb \in \Cbb^M$ accurately. Unfortunately, the resultant feedback overhead is too heavy even for a medium size of antenna ports at the BS, e.g., $M = 16$. Therefore, it calls for more effective design with light feedback. 
	\section{Proposed CSI Sensing Scheme} \label{sec:alg_design}
	
	This section delineates the design of our proposed CSI sensing scheme. The goal is to achieve the CSI sensing performance as good as the Type-II codebook feedback while using only PMI/CQI of the Type-I codebook.  Given the formulation in (\ref{eq:PR}) as a beginning, we propose three ingredients to achieve the desired goal,  namely parameter dimension reduction via spatial consistency,  hybrid precoder design, and a new constrained PR formulation.
	
	\subsection{Dimension Reduction} 

	As mentioned before, the number of measurements $T$ required to recover the CSI in PR (\ref{eq:PR}) is proportional to that of parameters $M$. In view of this, we try to utilize the low-rank structure of wireless channels to reduce the dimension of parameters. The flat fading DL channel vector can be expressed as \cite{DeepMIMO}
	\begin{equation} \label{eq:ch_model_dp}
		\begin{aligned}
			\hb = \sum_{\ell=1}^{n_L} \alpha_{\ell} \ab\left(\phi_{\mathrm{az}, \ell}, \phi_{\mathrm{el}, \ell}\right), 
		\end{aligned}
	\end{equation}
	which is composed of $n_L$ propagation paths. Each path is composed by a complex channel gain $\alpha_{\ell}$ and the array response vector $\ab(\phi_{\mathrm{az}, \ell}, \phi_{\mathrm{el}, \ell})$, where $\phi_{\mathrm{az}}$ and $\phi_{\mathrm{el}}$ correspond to the azimuth and elevation angles of departure (AoD), respectively.
	
	
	While $n_L$ is typically large, because of the limited local scattering effect in the propagation, most of the energy of the channel is dominated by a small number of propagation paths. Hence it is sufficient to have a good approximation of the CSI by only considering these dominant paths. 
	Specifically,  by taking the first $L \ll n_L$ dominant path components,  (\ref{eq:ch_model_dp}) can be approximated as
	\begin{equation}
		\hb \approx \Ab \alphab, \label{eq:h2g}
	\end{equation}
	where
	\begin{equation} \label{eq:D_path}
		\Ab =[\ab(\phi_{\mathrm{az}, 1}, \phi_{\mathrm{el}, 1}), \dots, \ab(\phi_{\mathrm{az}, L}, \phi_{\mathrm{el}, L})] \in \Cb^{M\times L} 
	\end{equation}
	and $\alphab = [\alpha_1, \dots, \alpha_L]^T \in 
	\Cbb^{L}$. 	
	Substituting (\ref{eq:h2g}) into (\ref{eq:PR}), the CSI sensing problem is updated as
	\begin{equation} \label{eq:PR_g:0}
		\min_{\gb \in \Cbb^L} \sum_{t=1}^T \big(q_t - | \ub_{j_t^*}^H  \Wb_t^H \Ab \gb|^2 \big)^2.
	\end{equation}
	
	Compared with (\ref{eq:PR}),  as long as  $L < M$, the CSI sensing problem (\ref{eq:PR_g:0}) can reduce the required rounds of feedback $T$ for achieving a good recovery performance.
	However, two fundamental questions regarding problem (\ref{eq:PR_g:0}) are 1) the construction of the basis matrix $\Ab$ and 2) the proper design of the precoder $\Wb_t$ for improved performance.  Next we show how they can be achieved by utilizing the spatial consistency of wireless channels.
	
	\subsection{Basis Construction by Spatial Consistency}
	The basis matrix $\Ab$ in (\ref{eq:D_path}) can be constructed by interpolating the associated path parameters of the TU from RUs' path parameters. The interpolation explicitly utilizes the spatial consistency. However, this construction approach requires the estimation of RUs' path parameters and well-designed path aligning algorithms before interpolation. The computational cost due to parameter estimation and path alignment at the BS cannot be neglected. 
	
	To avoid the problem, the basis matrix $\Ab$ is constructed by implicitly exploiting the spatial consistency instead. Specifically, given $n_R$ RUs near the TU, denote their corresponding CSI as $\hb_r \in \Cbb^M, r = 1, \dots, n_R$. Here we assume all UEs' locations are available at the BS so that the RUs can be selected from the neighborhood of the TU.  In practice, the locations of UEs can be attained by the global positioning system (GPS), 5G NR positioning \cite{TS_38455} or the combination of them,  and reported to the BS. 
	The CSI of RUs can be attained, for example, from the historical data stored at the BS or Type-II feedbacks.   By the spatial consistency, the AoDs of RUs should be close to those of the TU. In view of this, we assume that 
	\begin{equation}
		\hb \in \mathsf{Range} (\Hb)~ \text{and} ~\mathsf{Range} (\Hb) \approx \mathsf{Range} (\Ab),
	\end{equation} 
	where
	\begin{equation} \label{eq:H}
		\Hb = [\hb_1, \ldots, \hb_{n_R}].
	\end{equation}
	Therefore, $\Ab$ can be approximately constructed based on $\Hb$.
	Specifically, consider the singular value decomposition (SVD) of $\Hb$ as
	\begin{equation}\label{eq:C_h}
		\Hb = \sum_{i=1}^{\min\{M, n_R\}} \sigma_i \bar\vb_i \bar\ub_i^H		
	\end{equation}
	with $\sigma_1 \ge \dots \ge \sigma_{\min\{M, n_{R}\}}$ being the ordered singular values,  and the left and right singular vectors are $\{\bar \vb_i\}$ and $\{\bar \ub_i\}$, respectively.	 
	One may simply take
	\begin{equation} \label{eq:D_pca}
		\Db \triangleq [\bar\vb_1, \dots, \bar\vb_L].
	\end{equation}
	as an approximation of the basis of $\mathsf{Range}(\Ab)$. Correspondingly, (\ref{eq:h2g}) is replaced by
	\begin{equation}
		\hb \approx \Db \gb
	\end{equation}
	and the CSI sensing problem (\ref{eq:PR_g:0}) is updated as 
	\begin{equation} \label{eq:PR_g}
		\min_{\gb \in \Cbb^L} \sum_{t=1}^T \big( q_t - | \ub_{j_t^*}^H  \Wb_t^H \Db \gb |^2 \big)^2.
	\end{equation}
	It is worth noting that when only the Type-II codeword $\cb_r$ rather than $\hb_r$ of RU $r$ is available,  $\Hb$ in (\ref{eq:H}) can be replaced by
	\begin{equation} \label{eq:Cov_T2}
		\Hb = [\cb_1, \dots, \cb_{n_R}]
	\end{equation}  from the RUs.
	
	
	\subsection{Hybrid Precoding Design} \label{sec:hyb}
	The precoding matrix $\Wb_t$ at the BS is not only critical for the UE side channel estimation but also for the BS side CSI sensing.  The design of $\Wb_t$ is based on two important observations. First, $\Wb_t$ should be random which provides diversity and can improve the PR performance.  Specifically,  to learn about $\hb$ from as many different directions as possible, the precoding matrix applied at different time instants should be as diverse as possible \cite{Sidiropoulos_twc15}. Second, from (\ref{eq:CSI_eff}) $\Wb_t$ should be matched with $\hb$ to improve the received SNR at the TU side. 
	
	Based on the above two observations, we design $\Wb_t$ in a hybrid fashion, i.e.,
	\begin{equation} \label{eq:W_hyb}
		\Wb_t = \Wb_{t2} \Wb_{t1},
	\end{equation}
	where $\Wb_{t1} \in \Cbb^{M \times N_p}$ brings randomness in the sensing and $\Wb_{t2} \in \Cbb^{M \times M}$ is to improve the received SNR. The design of $\Wb_{t1}$ follows a similar idea in the PR literature, where the measurement vectors are usually sampled from a random Gaussian matrix \cite{PRIME_tsp16, Fuxiao_tsp18}. Thus, in our CSI sensing problem, $\Wb_{t1}$ is chosen to be a random Gaussion matrix, i.e.,
	\begin{equation} \label{eq:W_t1}
		\Wb_{t1}\sim \mathcal{CN}(\mathbf{0}, \sigma_w^2 \Ib),
	\end{equation}  
	where $\sigma_w$ is the standard deviation. 
	
	To improve the received SNR at the TU side, $\Wb_{t2}$ should concentrate on the main components of the CSI rather than the whole space. In other words, $\Wb_{t2}$ should match the space of $\hb$, i.e., $\textsf{Range}(\Db)$. Inspired by the idea, $\Wb_{t2}$ is chosen to be the projection operator on $\textsf{Range}(\Db)$ and is expressed as
	\begin{equation} \label{eq:proj}
		\Wb_{t2} = \Db \Db^H,
	\end{equation}
	where $\Db$ is given in (\ref{eq:D_pca}).
	
	It is not difficult to verify by checking (\ref{eq:W_hyb}) and (\ref{eq:CSI_eff}) that the hybrid design essentially projects the original channel to the estimated space of the principal components firstly and then applies the random Gaussian matrix $\Wb_{t1}$. The hybrid structure of $\Wb_t$ brings benefits to the reduction of feedback rounds $T$, especially for the wireless environment of copious propagation paths. Its efficacy will be shown in Section \ref{sec:simu}.
	\subsection{CPR Formulation}
	
	As described in Section \ref{sec:model}, the PMI fed back from the TU to the BS is selected by (\ref{eq:T1_sel}) with the Type-I codebook (\ref{eq:T1_cw}). As a result, the CSI to reconstruct should satisfy the following inequalities
	\begin{equation} \label{eq:T1_ineqs}
		|\hb^H \Wb_t \ub_{j}|^2 \le |\hb^H \Wb_t \ub_{j_t^*}|^2, \\ 
		\forall~ j \neq j_t^*,~t = 1, \dots, T.
	\end{equation}
	Notice that the inequalities in (\ref{eq:T1_ineqs}) actually specify a feasible region for the CSI $\hb$. Therefore,  incorporating them into problem (\ref{eq:PR_g}) would facilitate the search of the true CSI and enhance the sensing performance.
	By incorporating the inequalities in (\ref{eq:T1_ineqs}), problem (\ref{eq:PR_g}) is updated as
	\begin{subequations} \label{p:CPR}
		\begin{align}
			&\min_{\gb \in \Cbb^L} \quad  \sum_{t=1}^T \big(
			q_t - |\ub_{j_t^*}^H  \Wb_t^H \Db \gb|^2\big)^2\\
			&\label{p:CPR:cons}
			~\text{s.t.} \quad  |\ub_{j}^H  \Wb_t^H \Db\gb |^2 \le | \ub_{j_t^*}^H  \Wb_t^H \Db \gb |^2, 
			~ \forall j \neq j_t^*,  ~t = 1, \dots, T.
		\end{align}
	\end{subequations}

	There are three main challenges to solve the CPR problem (\ref{p:CPR}). First, the objective in (\ref{p:CPR}) is non-convex. Second, the constraints are also non-convex and difficult to project onto.  As a result, the projected gradient descent method popularly used in the literature for solving the PR problem are no longer suitable for problem (\ref{p:CPR}). Last but not least, the number of constraints in (\ref{p:CPR:cons}) is $T(n_{T1}-1)$, which is very large due to the codebook size $n_{T1}$. For instance, $n_{T1} = 512$ for the Type-I codebook of $N_p = 16$ CSI ports, as specified by 3GPP \cite{TS_38214}. Given a small value of $T$ like $T = 2$, the total number of constraints in problem (\ref{p:CPR}) can easily exceed one thousand, which poses a big challenge for efficient CSI recovery.
	
	%
	%
	%
	

	\section{Efficient Algorithm Designs} \label{sec:MECS_SGDA}
	
	In this section, two algorithms to solve the CPR problem (\ref{p:CPR}) are developed.   We first review the algorithm in \cite{spawc22} that is based on the PD optimization and eigenvalue decomposition (EVD). Then, to overcome the challenge of a large number of constraints in (\ref{p:CPR}),  we further propose a computationally efficient two-stage method that first removes redundant constraints from (\ref{p:CPR}) followed by solving the Lagrange dual problem of (\ref{p:CPR})  via a simple first-order algorithm.
	%
	%
	
	\subsection{PD-EVD Algorithm}
	
	By defining $\mub_{j_t} = \Db^H \Wb_t \ub_{j}$ and
	introducing matrices $\Ub_{j_t} = \mub_{j_t} \mub_{j_t}^H$ and $\Gb = \gb \gb^H$, problem (\ref{p:CPR}) is equivalently expressed as
	\begin{subequations} \label{p:CPR_g_SDR}
		\begin{align}
			\min_{\Gb, \gb} &\quad   \sum_{t=1}^T \big(
			q_t - \text{Tr}(\Ub_{j_t^*} \Gb) \big)^2 \\
			\text{s.t.} & \quad 
			\text{Tr}(\Ub_{j_t} \Gb) \le \text{Tr}(\Ub_{j_t^*} \Gb), 
			~\forall j_t \neq j_t^*, t = 1, \dots, T,\\
			& \quad \Gb = \gb \gb^H. \label{c2:p:CPR_g_SDR}
		\end{align}
	\end{subequations}
	One can observe that the objective and all the constraints in problem (\ref{p:CPR_g_SDR}) are convex except (\ref{c2:p:CPR_g_SDR}).  
	One alternative to handling \eqref{p:CPR_g_SDR} is to adopt the well-known semidefinite relaxation (SDR) technique by simply relaxing $ \Gb = \gb \gb^H $ to $ \Gb \succeq \zerob$ (positive semidefinite).  
	However,  this approach has two drawbacks. 
	First, the rank-one condition in (\ref{c2:p:CPR_g_SDR}) is not guaranteed. Moreover, the resultant relaxation problem would still be computationally expensive due to the huge number of constraints. To cope with these problems, we instead develop a customized algorithm for problem (\ref{p:CPR_g_SDR}) based on the primal-dual optimization.
	
	By introducing dual variables $\lambda_{j_t} \ge 0, j_t = 1, \dots, n_{T1}, j_t \neq j_t^*, t = 1, \dots, T$, the Lagrangian of (\ref{p:CPR_g_SDR}) can be written as
	\begin{equation}
		\begin{aligned}
			\Lc(\Gb, \lambda_{j_t}) & =  \sum_{t=1}^T \big( q_t - \text{Tr}(\Ub_{j_t^*} \Gb) \big)^2 + \sum_{t=1}^T \sum_{j_t \neq j_t^*} \lambda_{j_t} \big(\text{Tr}(\Ub_{j_t} \Gb) -\text{Tr}(\Ub_{j_t^*} \Gb) \big),\\
			& =  \sum_{t=1}^T \big( q_t - \text{Tr}(\Ub_{j_t^*} \Gb) \big)^2 + \sum_{t=1}^T \sum_{j_t \neq j_t^*}\lambda_{j_t}  \text{Tr}(\Vb_{j_t} \Gb), \\
			& = f_1(\Gb) + f_2(\Gb, \lambda_{j_t}),
		\end{aligned}
	\end{equation}
	with $\Vb_{j_t} = \Ub_{j_t}  - \Ub_{j_t^*}$.
	Then the dual problem of (\ref{p:CPR_g_SDR}) is given by
	\begin{equation} \label{p:dual_opt}
		\max_{ \substack{ \lambda_{j_t}\ge 0, \forall j_t \neq j_t^* \\ t = 1, \dots, T }} ~ \min_{\Gb = \gb\gb^H} \quad \Lc(\Gb, \lambda_{j_t}).
	\end{equation}
	An algorithm to solve (\ref{p:dual_opt}) can be developed by solving the inner problem and the outer problem alternatingly until a predefined stopping condition is satisfied.
	In our design,  the inner problem of (\ref{p:dual_opt}) is further processed by the majorization-minimization method following the PRIME algorithm in \cite{PRIME_tsp16} to obtain a closed-form solution in each iteration, while the outer problem is solved by the projected gradient ascent method.  
	
	
	Specifically, given $\lambda_{j_t}$ in problem (\ref{p:dual_opt}), the inner problem is 
	\begin{subequations} \label{p:inner}
		\begin{align}
			\min_{\Gb, \gb} \quad & f_1(\Gb) + f_2(\Gb, \lambda_{j_t}) \\ 
			\text{s.t.} \quad & \Gb = \gb\gb^H.
		\end{align}
	\end{subequations}
	To achieve efficient computation, the first term $f_1(\Gb)$ is further majorized at the point $\Gb^{(k)}$ as 
	\begin{equation}
		\tilde{f}_1(\Gb,\! \Gb^{(k)}) \!=\! \beta \text{Tr}(\Gb\Gb)\! +\! 2\sum_{t=1}^T \!\text{Tr}(\Gb \Ub_{j_t^*})  \text{Tr}(\Gb^{(k)} \!\Ub_{j_t^*}) \!- \! 2\beta \text{Tr}(\Gb \Gb^{(k)}) \!-\! 2\sum_{t=1}^T\! q_t \text{Tr}(\Ub_{j_t^*}\Gb),
	\end{equation}
	with $1/\beta \le 1/\lambda_{max}(\Qb)$ to be a step size parameter, where $\lambda_{max}(\Qb)$ takes the largest eigenvalue of $\Qb = \sum_{t=1}^T \text{vec}(\Ub_{j_t^*})\text{vec}(\Ub_{j_t^*})^T$. Therefore, the majorization problem for (\ref{p:inner}) can be written as 
	\begin{subequations} \label{p:inner_major}
		\begin{align}
			\min_{\Gb, \gb} \quad & \tilde{f}_1(\Gb, \Gb^{(k)})  + f_2(\Gb, \lambda_{j_t}) \\
			\text{s.t.} \quad & \Gb = \gb\gb^H.
		\end{align}
	\end{subequations}
	It is not difficult to show that problem (\ref{p:inner_major}) is equivalent to 
	\begin{equation} \label{p:leadingEig}
		\underset{\gb}{\min}\left\|\gb \gb^{H}- \Rb^{(k)} \right\|_{F}^{2},
	\end{equation}
	where  
	\begin{equation} \label{expression:R}
		\begin{aligned}
			\Rb^{(k)} =  \gb^{(k)}\left(\gb^{(k)}\right)^{H}+\frac{1}{\beta} \Mb \operatorname{diag}(\{ q_t- \mub_{j_t^*}^H\gb^{(k)} \}_{t=1}^T) \Mb^{H}  -\frac{1}{\beta} \sum_{t=1}^T \sum_{j_t \neq j_t^*}\lambda_{j_t}  \Vb_{j_t}, 
		\end{aligned}
	\end{equation}
	$\Mb = [\ub_{j_1}^*, \dots, \ub_{j_T}^*] \in \Cbb^{L\times T}$ and $\text{diag}(\{a_i\}_{i=1}^T)$ is a diagonal matrix whose diagonal entries are $a_1, \dots, a_T$.
	Obviously, problem (\ref{p:leadingEig}) is a leading eigenvector problem admitting a closed-form solution
	\begin{equation} \label{expression:g}
		\gb^{(k),*} = \sqrt{\lambda_{max} (\Rb^{(k)})} \ub_{max} (\Rb^{(k)}),
	\end{equation}
	where $\lambda_{max}$ and $ \ub_{max}$ represents the leading eigenvalue and eigenvector of $\Rb^{(k)}$ and can be computed via the power method efficiently. 
	
	Once the inner problem is solved, the dual variables in the outer problem of (\ref{p:dual_opt}) can be updated via the projected dual ascent method, i.e.
	\begin{equation} \label{expression:lambda}
		\lambda_{j_t} = \max[\lambda_{j_t} + \gamma \text{Tr}(\Vb_{j_t} \Gb^*), 0],
	\end{equation}
	where $\gamma > 0$ is the step size. The steps of the overall algorithm are summarized in Algorithm \ref{alg:prime_T1}.
	
	\setlength{\textfloatsep}{5pt}
	\begin{algorithm}[t]
		\caption{PD-EVD Algorithm}
		\begin{algorithmic}[1] \label{alg:prime_T1}
			\STATE {\bf Given} $\Wb_t, q_t, \Db$ and $\ub_{j_t^*}, t = 1, \dots, T$, construct $\Mb, \Ub_{j_t}$ and $ \Vb_{j_t}$.
			\STATE Initialize dual variables $\lambda_{j_t} \leftarrow 0, j_t = 1, \dots, n_{T1}, j_t \neq j_t^*, t = 1, \dots, T$. Initialize $\gb^{(0)} $ by $ \gb^{(0)} \leftarrow \ub_{max}(\Mb \text{diag}( \{q_t\}_{t=1}^T) \Mb^H)$.
			\REPEAT
			\STATE Initialize $k = 0$.
			\REPEAT
			\STATE Compute $\Rb^{(k)}$ according to (\ref{expression:R}).
			\STATE Compute $\gb^{(k), *}$ according to (\ref{expression:g}).
			\STATE Update $k \leftarrow k +1$, $\gb^{(k)} \leftarrow \gb^{(k), *}$. 
			\UNTIL a predefined stopping criterion is satisfied.
			\STATE Update $\lambda_{j_t}$ by (\ref{expression:lambda}).
			\UNTIL a predefined stopping criterion is satisfied.
			\STATE {\bf Output} $\gb^*$, and reconstruct the CSI by $\hb^* = \Db \gb^*$.
		\end{algorithmic}
	\end{algorithm}
	
	Experiment results presented in Section \ref{sec:simu} will show that the above PD-EVD algorithm can provide promising CSI  sensing performance with a significantly reduced feedback overhead.  Nevertheless, it is also found that the computation of PD-EVD is still heavy.  One reason is that step 7 in PD-EVD involves the computation of the leading eigenvector that is not computationally cheap.  In addition,  PD-EVD needs to process all the constraints of problem (\ref{p:dual_opt}), which is another bottleneck to achieve efficient solution. 
	In view of this,  we  next develop a two-stage approach to solve the CPR problem (\ref{p:CPR}) via the MECS construction and the first-order SGDA algorithm.
	%
	\subsection{Two-Stage MECS-SGDA Algorithm}
	
	With the observation that many of the constraints in the CPR problem (\ref{p:CPR}) may not be active at the optimum,  finding the effective ones and removing those redundant ones from the problem would achieve complexity reduction.  On the other hand, to {avoid computing the leading eigenvector}, we consider the first-order methods to handle problem (\ref{p:CPR}). 
	This idea gives rise to a two-stage method as described below.	
	%
	
	{\bf MECS Construction:} The task of constructing the MECS $\mathcal{S}$ is to find the minimal number of constraints that effectively bound the true $\gb$ followed by eliminating those redundant constraints from problem (\ref{p:CPR}). However, it is highly non-trivial to find all the effective constraints in (\ref{p:CPR:cons}) theoretically. Instead,  we resort to a suboptimal method. 
	The construction process starts with an initial point as well as an empty MECS, and adding the unsatisfied constraints to the MECS gradually until a point feasible to all the constraints is found.  
	
	Specifically, initialize $\mathcal{S} \triangleq \emptyset$ and the iteration index $k=0$. Given an initial point $\gb^{k}$, these unsatisfied constraints at $\gb^{k}$ constitute a set $\mathcal{S}^{k}$ and the MECS $\mathcal{S}$ is updated by 
	\begin{equation}
		\mathcal{S} = \mathcal{S} \cup \mathcal{S}^{(k)}.
	\end{equation}
	Afterwards, $\gb^{k}$ is updated by solving the following feasibility problem	
	\begin{subequations} \label{p:fea}
		\begin{align}
			\text{find}~ & \gb \\
			\text{s.t.}~& \gb^H \Vb_{j_t} \gb \le 0, j_t \in \mathcal{S},
		\end{align}
	\end{subequations}	
	where $\Vb_{j_t} \triangleq \Ub_{j_t} - \Ub_{j_t^*}$. The updates of $\mathcal{S}$ and $\gb$ proceed alternatingly until a point meeting all the constraints in  (\ref{p:CPR:cons}) is found. The details of the MECS construction are given in Algorithm \ref{alg:MECSC}. Because the original CSI satisfies all the constraints in (\ref{p:CPR:cons}) and the set $\mathcal{S}$ will not shrink during the iterations, it is not difficult to verify that Algorithm \ref{alg:MECSC} will terminate within a finite number of iterations.
	%
	
	\begin{algorithm}[t]
		\caption{Stage I - MECS Construction}
		\begin{algorithmic}[1] \label{alg:MECSC}
			\STATE {\bf Given} $\Wb_t, q_t, \Db$ and $\ub_{j_t^*}, t = 1, \dots, T$, construct $\Mb$.
			\STATE Initialize the MECS as $\mathcal{S} \leftarrow \emptyset$. {Initialize} $k = 0$ and $ \gb^{(0)} \leftarrow \ub_{max}(\Mb \text{diag}(\{q_t\}_{t=1}^T) \Mb^H)$. 
			\REPEAT
			\STATE Substitute $\gb^{(k)}$ into (\ref{p:CPR:cons}) to get the set  $\mathcal{S}^{(k)}$ defined by these unsatisfied inequalities.
			\STATE Update $\mathcal{S} = \mathcal{S} \cup \mathcal{S}^{(k)}$.
			\STATE Solve problem (\ref{p:fea}) defined by the constraint set $\mathcal{S}$ to get a feasible point $\gb^{*}$.
			\STATE $k \leftarrow k +1, \gb^{(k)} \leftarrow \gb^{*}$.
			\UNTIL $\mathcal{S}^{(k)} = \emptyset$.
			\STATE {\bf Output} $\gb^{(k)}$ and $\mathcal{S}$.
		\end{algorithmic}
	\end{algorithm}

	The main step in Algorithm \ref{alg:MECSC} is to solve problem (\ref{p:fea}), which is a non-convex quadratically constrained problem. 
	While the consensus-ADMM algorithm proposed in \cite{HKJ} can undertake the task, the approach involves eigen-decomposition that is usually not computationally cheap.  
	Therefore, problem (\ref{p:fea}) is handled in a different way in our work.
	Specifically, by introducing a continuous and non-convex function	
	\begin{equation}
		f_{j_t}(\gb) \triangleq \left(\max \{\gb^H \Vb_{j_t} \gb, 0\} \right) ^2,	
	\end{equation} 
	problem (\ref{p:fea}) can be transformed into
	\begin{equation} \label{p:proximal}
		\min_{\gb} ~ \sum_{j_t \in \mathcal{S}} f_{j_t}(\gb).	
	\end{equation}
	Thus, problem (\ref{p:proximal}) becomes an unconstrained problem with a smooth objective function and can be efficiently solved by the gradient descent method or ADMM-based methods \cite{proximal_ADMM}. 
	
	{\bf SGDA Algorithm:} By Algorithm \ref{alg:MECSC}, a feasible point satisfying all constraints of problem \eqref{p:CPR} and a MECS $\mathcal{S}$ can be obtained.  Then,  what remains is to solve problem \eqref{p:CPR} with the MECS:
	\begin{subequations} \label{p:orig}
		\begin{align}
			\min_{\gb \in \Cbb^L} ~& \sum_{t=1}^T q_t^2 - 2 q_t \gb^H \Ub_{j_t^*} \gb + (\gb^H \Ub_{j_t^*} \gb)^2,\\
			\text{s.t.} ~& \gb^H \Ub_{j_t} \gb \le \gb^H \Ub_{j_t^*} \gb, ~ j_t \in \mathcal{S}.
		\end{align}
	\end{subequations}
	Rather than considering the PD-EVD algorithm or other methods based on the convex-concave procedure (CCP) \cite{park2017general},  we propose to employ first-order methods to handle problem \eqref{p:orig}.  
	%
	%
	%
	In particular,  inspired by the smoothed gradient descent ascent (SGDA) algorithm framework proposed in \cite{ZJW},  we consider the Lagrangian dual problem of (\ref{p:orig}) as follows 
	\begin{equation} \label{p:sgda:1}
		\max_{\{ \nu_{j_t}  \ge 0\}}~\min_{\gb \in \Cbb^L}~ \mathcal{L}(\gb, \nu_{j_t}),
	\end{equation} 
	where
	\begin{equation}
		\begin{aligned}
			\mathcal{L}(\gb, \nu_{j_t}) = & \sum_{t=1}^T \big( q_t^2 - 2 q_t \gb^H \Ub_{j_t^*} \gb  + (\gb^H \Ub_{j_t^*} \gb)^2 \big)  + \sum_{j_t \in \mathcal{S}} \nu_{j_t} (\gb^H \Ub_{j_t} \gb - \gb^H \Ub_{j_t^*} \gb),
		\end{aligned}
	\end{equation} is the Lagrangian function and
	$\nu_{j_t} \ge 0$ are dual variables.
	
	Note that \eqref{p:sgda:1} is a minmax problem which is concave w.r.t.  $\{\nu_{j_t}\}$ but not necessarily convex w.r.t.  $\gb$.  To overcome the non-convexity,  
	SGDA introduces an auxiliary variable $\zb$ and solves the following problem as
	\begin{equation}\label{p:sgda:2}
		\max_{\nu_{j_t} \ge 0}~\min_{\gb \in \Cbb^L}~ \hat{\mathcal{L}}(\gb, \nu_{j_t}, \zb) \triangleq \mathcal{L}(\gb, \nu_{j_t}) + \frac{p}{2} \|\gb - \zb\|^2,
	\end{equation}
	with $p > 0$ as a constant parameter and is large enough so that $\hat{\mathcal{L}}(\gb, \nu_{j_t}, \zb)$ is convex w.r.t. $\gb$.
	In SGDA, the gradient descent and ascent are conducted in the alternating fashion w.r.t. the primal variable $\gb$ and dual variables $\{\nu_{j_t}\}$,  while the auxiliary $\zb$ is updated via an averaging step.
	The details of SGDA for problem (\ref{p:sgda:1}) are summarized in Algorithm \ref{alg:GDA},
	where $\beta, s_1, s_2$ are stepsize parameters. 
	As seen,  the algorithm involves only simple gradient descent/ascent updates and therefore are computationally cheap in each iteration.  
	Conditions for the SGDA algorithm to converge to a stationary solution of problem \eqref{p:sgda:1} have been proved in  \cite{ZJW}.
	
	{Notice that Algorithm \ref{alg:GDA} is applied to solve problem (\ref{p:sgda:1}) rather than the original problem (\ref{p:orig}). Due to the non-convexity of problem (\ref{p:orig}), Algorithm \ref{alg:GDA} cannot guarantee to find a solution satisfying all the constraints in (\ref{p:orig}). It is necessary to emphasize that the goal here is not to meet all the constraints but to find a solution satisfying most of the constraints in order to improve the sensing performance of the unconstrained counterpart. Moreover, the tests in Section \ref{sec:simu} showcase that the solution returned by Algorithm \ref{alg:GDA} indeed meets most of the constraints and achieves a noticeable improvement on the sensing performance.}

	\setlength{\textfloatsep}{3pt}
	\begin{algorithm}[t]
		\caption{Stage II - SGDA for Problem (\ref{p:sgda:1})}
		\begin{algorithmic}[1] \label{alg:GDA}
			\STATE {\bf Initialize} $k = 0, \gb^0, \zb^0, \nu_{j_t}^0$ and $\beta \in (0, 1]$. Step size $s_i > 0, i = 1, 2.$
			\REPEAT
			\STATE $\gb^{k+1} \leftarrow \gb^k - s_1 \nabla_{\gb} \hat{\mathcal{L}}(\gb, \nu_{j_t}; \zb) $.
			\STATE $\nu_{j_t}^{k+1} \leftarrow [\nu_{j_t}^k + s_2 \nabla_{\nu_{j_t}} \hat{\mathcal{L}}(\gb, \nu_{j_t}; \zb) ]_{+}$.
			\STATE $\zb^{k+1} \leftarrow \zb^{k} + \beta (\gb^{k+1} - \zb^{k})$.
			\UNTIL {a predefined stopping criterion is satisfied}
			\STATE {\bf Output} $\gb^{(k)}$.
		\end{algorithmic}
	\end{algorithm}
	
	\section{Extension to Multi-carrier Systems} \label{sec:mulC}
	So far,  the proposed CSI sensing scheme considers only the flat fading channels.  For frequency selective fading channels,  we consider the multi-carrier transmission system (e.g., OFDM) and extend the CSI sensing scheme to this scenario.  Rather than applying it to each of the subcarriers,  we explore the channel correlation between different subcarriers.

	Assume that the multi-carrier system has $n_C$ subcarriers. The DL channel in the $k$-th subcarrier can be expressed as \cite{DeepMIMO}	
	\begin{equation} \label{eq:ch_mulC}
		\begin{aligned}
			\hb^{(k)} = &\sum_{\ell=1}^{n_L} \sqrt{\frac{\rho_{\ell}}{n_C}} e^{j (\theta_{\ell} - \frac{2\pi k}{n_C} \tau_{\ell} B)} \ab\left(\phi_{\mathrm{az}, \ell}, \phi_{\mathrm{el}, \ell}\right),  
		\end{aligned}	
	\end{equation}
	$ k = 1, \dots, n_C,$	where $\rho_{\ell}, \theta_{\ell}$ and $\tau_{\ell}$ are the power, phase and delay of the $\ell$-th propagation path, respectively.
	Correspondingly, the CSI to sense is 	
	\begin{equation} \label{eq:csi_mc_f}
		\Hb = [\hb^{(1)}, \dots, \hb^{(k)}, \dots, \hb^{(n_C)}] \in \Cbb^{M \times n_C}.	
	\end{equation}
	
	Since the number of propagation paths $n_L$ is usually much smaller than the number of subcarriers $n_C$,  we consider transforming $\Hb$ into the antenna-delay domain by	
	\begin{equation} \label{eq:csi_mt}
		\widetilde{\Hb} = \Hb \Fb,	
	\end{equation}
	where	$\Fb \in \mathbb{C}^{n_C \times \tilde n_L}$ contains the first $ \tilde n_L$ columns of the inverse discrete Fourier transform (IDFT) matrix of size $n_C$.
	Similar to (\ref{eq:h2g}), $\widetilde{\Hb}$ can be further approximated by the first few dominant paths (say $\tilde L$ paths), like
	\begin{equation} \label{eq:HDG}
		\widetilde{\Hb} \approx \widetilde{\Ab} \Gb,
	\end{equation}	
	where the basis matrix $\widetilde{\Ab} \in \Cbb^{M \times \tilde{L}}$ and the coefficient matrix $\Gb \in \Cbb^{\tilde{L} \times \tilde{n}_L}$.  
	
	
	From the aforementioned derivations, one can observe that the number of the parameters to recover
	is reduced from $M n_C$ to $\tilde{L} \tilde{n}_L$.  The transformation from $\Hb$ to $\Gb$ in essence utilizes the correlation of the CSI on the angular and frequency domains jointly, making it possible to further cut down the feedback overhead via sensing the dimension reduced $\Gb$.
	Correspondingly, the extensions of the basis matrix construction, precoder design and CPR formulation to the multi-carrier system are developed as follows. 
	
	{\bf Basis matrix construction:} Notice that $\widetilde{\Ab}$ is the basis matrix for the CSI in the delay domain. 
	Denote the CSI of RUs as $\Hb_{r} \!\in \!\mathbb{C}^{M\times n_C}, r = 1, \dots, n_R$.\! We take their delay-domain counterparts by	
	\begin{equation}
		\widetilde{\Hb}_{r} = \Hb_{r} \Fb \in \Cbb^{M \times \tilde n_L}, ~r = 1, \dots, n_R.	
	\end{equation}
	Similar to (\ref{eq:C_h}) based on the spatial consistency,  by taking the $\tilde L$ left principal singular vectors of 
	$[\widetilde{\Hb}_{1}, \dots, \widetilde{\Hb}_{n_R}]$,
	we can obtain a basis matrix for $\mathsf{Range}(\widetilde \Ab)$,  denoted by $\widetilde\Db \in \Cbb^{M \times \tilde{L}}.$
	
	%
	
	{\bf Precoder design:} Define $\Wb_{t}^k$ as the precoding matrix for each subcarrier $k$ in the $t$-th round.  Like (\ref{eq:W_hyb}), $\Wb_{t}^k$ is constructed in a hybrid mode as	
	\begin{equation} \label{eq:W_tk}
		\Wb_{t}^k = \Wb_{t2}^k \Wb_{t1}, 	
	\end{equation}	 
	where $\Wb_{t1}$ is still sampled from the complex Gaussian distribution, i.e., $\Wb_{t1} \sim \mathcal{CN}(\mathbf{0}, \sigma_w^2 \Ib)$. The construction of $\Wb_{t2}^k$ relies on the basis matrix for the subspace that $\hb^{(k)}$ lie in.  Analogously,  we use the CSI of RUs	
	\begin{equation} \label{eq:csi_ru_mc subcarrier k}
		\Hb^{(k)} = [\hb_1^{(k)}, \dots, \hb_{n_R}^{(k)}]	
	\end{equation}
	and take the first $\tilde{L}$ principal left singular vectors of $\Hb^{(k)}$,  denoted by $\Db^{(k)}  \in \mathbb{C}^{M\times \tilde L}$,  as the basis matrix.  Then,  
	%
	the precoder $\Wb_{t2}^k$ is chosen to be 
	\begin{equation}
		\Wb_{t2}^k = \Db^{(k)} (\Db^{(k)})^H,	
	\end{equation}
	in order to maximize the received SNR at each subcarrier $k$.
	
	{\bf CPR formulation:} With the precoder $\Wb_{t}^k$ over the $k$-th sub-carrier, the effective CSI estimated at the TU is	 
	\begin{equation} \label{eq:CSI_eff_k}
		\hat\hb_{t}^{(k)} = (\Wb_{t}^k)^H \hb^{(k)} + \eb_{t}^{(k)},	
	\end{equation}
	where $\eb_{t}^{(k)}$ is the channel estimation error.
	Like (\ref{eq:T1_sel}), the PMI from the Type-I codebook is selected via	
	\begin{equation} \label{eq:T1_sel_k}
		j_{tk}^* = \arg \max_{j = 1, \dots, n_{T1}} |\ub_{j}^H \hat\hb_{t}^{(k)}|, ~t = 1, \dots, T,	
	\end{equation} 
	and the corresponding CQI is computed as	
	\begin{equation} \label{eq:CQI_k}
		q_{tk} = | \ub_{j_{tk}^*}^H \hat\hb_{t}^{(k)}|^2,~ t = 1, \dots, T. 	
	\end{equation} 
	As a result,  similar to (\ref{eq:T1_ineqs}), the associated constraints indicated by the PMI can be expressed as	
	\begin{equation} \label{eq:T1_ineqs_mc}
		\begin{aligned}
			&|(\hb^{(k)} )^H \Wb_{t}^k \ub_{j}|^2 \le |(\hb^{(k)} )^H \Wb_{t}^k \ub_{j_{tk}^*}|^2,~  
			\forall j \neq j_{tk}^*, t = 1, \dots, T.
		\end{aligned}		
	\end{equation}
	With the above basis matrix $\widetilde\Db$, we approximately have \footnote{In general, $\Hb = \widetilde{\Hb}\Fb^H$ does not hold as $\Fb$ is the truncated IDFT matrix and $\Fb \Fb^H \neq \Ib_{n_C}$. The rationale behind the approximation is that the energy of the CSI mainly dominate parts of the temporal components because of the limited local scattering effect. That is, given $[\widetilde{\Hb}, \Hb_{res}] \triangleq \Hb \Fb_0$ as the CSI in the time domain with $\Fb_0 \in \Cb^{n_C \times n_C}$ being the full IDFT matrix, the absolute values of the entries in $\Hb_{res}\in\Cbb^{n_c \times (n_c - \tilde{n}_L)}$ are almost zero. Therefore, $\Hb$ can be approximately recovered by $\widetilde{\Hb} \Fb^H$.  }	
	\begin{equation}
		\hb^{(k)} \approx \widetilde \Hb {\fb}^{(k)} 
		\approx \widetilde\Db\Gb {\fb}^{(k)} ,	
	\end{equation}
	where ${\fb}^{(k)}$ is the $k$-th column vector of $\Fb^H$. The CPR problem for the multi-carrier system is formulated as
	\begin{subequations} \label{p:CPR_mulC}
		\begin{align}
			\min_{\Gb \in \Cbb^{\tilde{L} \times \tilde{n}_L}} & \sum_{t =1}^T \sum_{k=1}^{n_C} \big(q_{tk} - \big|(\hb^{(k)} )^H \Wb_{t}^k \ub_{j_{tk}^*} \big|^2 \big)^2\\
			\text{s.t.} &~ (\ref{eq:T1_ineqs_mc}),~ k = 1, \dots, n_C,\\
			& ~\hb^{(k)} = \widetilde{\Db}\Gb {\fb}^{(k)},~ k = 1, \dots, n_C.
		\end{align}
	\end{subequations} 
	Notice that problem (\ref{p:CPR_mulC}) can also be solved by the proposed two-stage MECS-SGDA algorithm.  
	
	Until now, the extension to the multi-carrier system is based on the setting that the PMI and CQI are fed back per carrier. 
	For more flexible resource utilization, the computation and feedback of PMI and CQI  can be conducted over a group of sub-carriers like physical resource block (PRB) rather than per carrier.  For that scenario, we have the following remark. 
	\begin{Rmk} \label{remark:mulC}{\rm
			Assume that the overall frequency band is divided into $n_g$ groups of carriers, and each group contains $n_c$ subcarriers. Denote the estimate of the effective channel over the $k$-th carrier within the $b$-th group as $\hat\hb_{tb}^{(k)}, k = 1, \dots, n_c, b = 1, \dots, n_g, t = 1, \dots, T$. The PMI in the $t$-th round for the $b$-th group can be selected by
			\begin{equation}
				j_{tb}^*  = \arg \max_{j \in \{1, \dots, n_{T1}\} } \sum_{k = 1}^{n_c} |{\ub}_{j}^H \hat\hb_{tb}^{(k)}|^2,
			\end{equation}
			and the CQI can be computed as 
			\begin{equation}
				q_{tb} = \sum_{k = 1}^{n_c} |{\ub}_{j_{tb}^*}^H \hat\hb_{tb}^{(k)}|^2.
			\end{equation}
			It is not difficult to extend our previous designs to the scenario, and our proposed algorithms are still applicable.}
	\end{Rmk}
	\section{Simulation Results \label{sec:simu}}
	
	In this section, the proposed CSI sensing scheme is evaluated extensively based on the DeepMIMO dataset \cite{DeepMIMO} and the CSI generated by QuaDriGa platform\footnote{https://quadriga-channel-model.de/}. 
	Two metrics are applied to evaluate the performance of the sensed CSI $\hb^*$ -- the correlation 
	\begin{equation}
		\rho(\hb, \hb^*) = \frac{|\hb^H \hb^*|}{\|\hb\|_2 \|\hb^*\|_2},
	\end{equation}
	and the normalized mean squared error with a common phase rotation ($\text{NMSE}_r$) \cite{PRIME_tsp16} defined as
	\begin{equation}
		\text{NMSE}_r(\hb, \hb^*) = \min_{\psi \in [0, 2 \pi]}\frac{\| \hb - e^{j\psi}\hb^* \|_2}{\| \hb  \|_2}. 
	\end{equation}
	
	\subsection{Performance on DeepMIMO Dataset}
	The test data is from the DL CSI from BS 4 to user $63\sim149$ within row $810\sim 963$ in the DeepMIMO `O1' scenario. The DL CSI consists of $13\sim 16$ paths in that area, which is almost the most complicated propagation area covered by BS 4. To examine the performance gain of our proposed scheme clearly, $n_{TU} = 2000$ samples are selected from the total $n_s = 13398$ samples in the area. Besides, $20\% n_s$ other UE are selected to model a given dataset of RUs whose Type-II codewords are available. The other parameters in experiments are given in Table \ref{Tb:simu_DM}. For each TU, $n_R = 10$ RUs with the smallest distance to the TU are selected from the RU dataset. The Type-II codewords of RUs are used to construct $\Hb$ as (\ref{eq:Cov_T2}). Unless otherwise specified, the number of CSI ports is set as $N_p = M$. The value $\sigma_w^2$ to generate the random $\Wb_{t1}$ in (\ref{eq:W_t1}) is set to 1. For each TU, $n_{GS} = 10$ $\Wb_{t1}$s are randomly generated and the performance of the sensed CSI is averaged over them. The presented performance is averaged over all TUs.
	
	\begin{table}[]
		\centering
		\caption{The parameters of DeepMIMO dataset.}
		\begin{tabular}{l|llllll}   \hline
			Parameter & Value \\ \hline
			BS height &  6 meter\\
			UE height & 1.5 meter \\
			Number of antennas at the BS, $M = N_{y} N_{z}$ & $32$  \\
			Number of antennas in the horizontal dimension, $N_{y}$ & 8 \\ 
			Number of antennas in the vertical dimension, $N_{z}$ & 4 \\
			Antenna spcaing & $0.5 \lambda$ \\
			Center carrier frequency $f_c$ (GHz) & 3.5\\ 
			Oversampling rate of the Type-I, Type-II codebook $[O_1, O_2]$ & $[4, 4]$  \\
			\hline
		\end{tabular}
		\label{Tb:simu_DM}
		\vspace{0.3cm}
	\end{table}
	
	\subsubsection{\bf Significance of dimension reduction}
	The basis matrix in (\ref{eq:D_pca}) designed by exploiting the spatial consistency over the CSI of nearby UEs plays a critical role in the dimension reduction. To show that, two schemes are compared. The first one solves problem (\ref{eq:PR}) to obtain $\hb^* \in \Cbb^{32}$ without dimension reduction, while the second one solves problem (\ref{eq:PR_g}) with $\gb \in \Cbb^5$. The basis matrix $\Db$ is constructed according to (\ref{eq:Cov_T2}) in the second scheme. Both problem (\ref{eq:PR}) and problem (\ref{eq:PR_g}) are solved by the PRIME-power-acce algorithm (PRIME for short) \cite{PRIME_tsp16}. In the test, the BS only applies the random precoding $\Wb_t \triangleq \Wb_{t1}$.
	
	\begin{figure}[t] 
		\centering	
		{\includegraphics[width=0.85\textwidth]{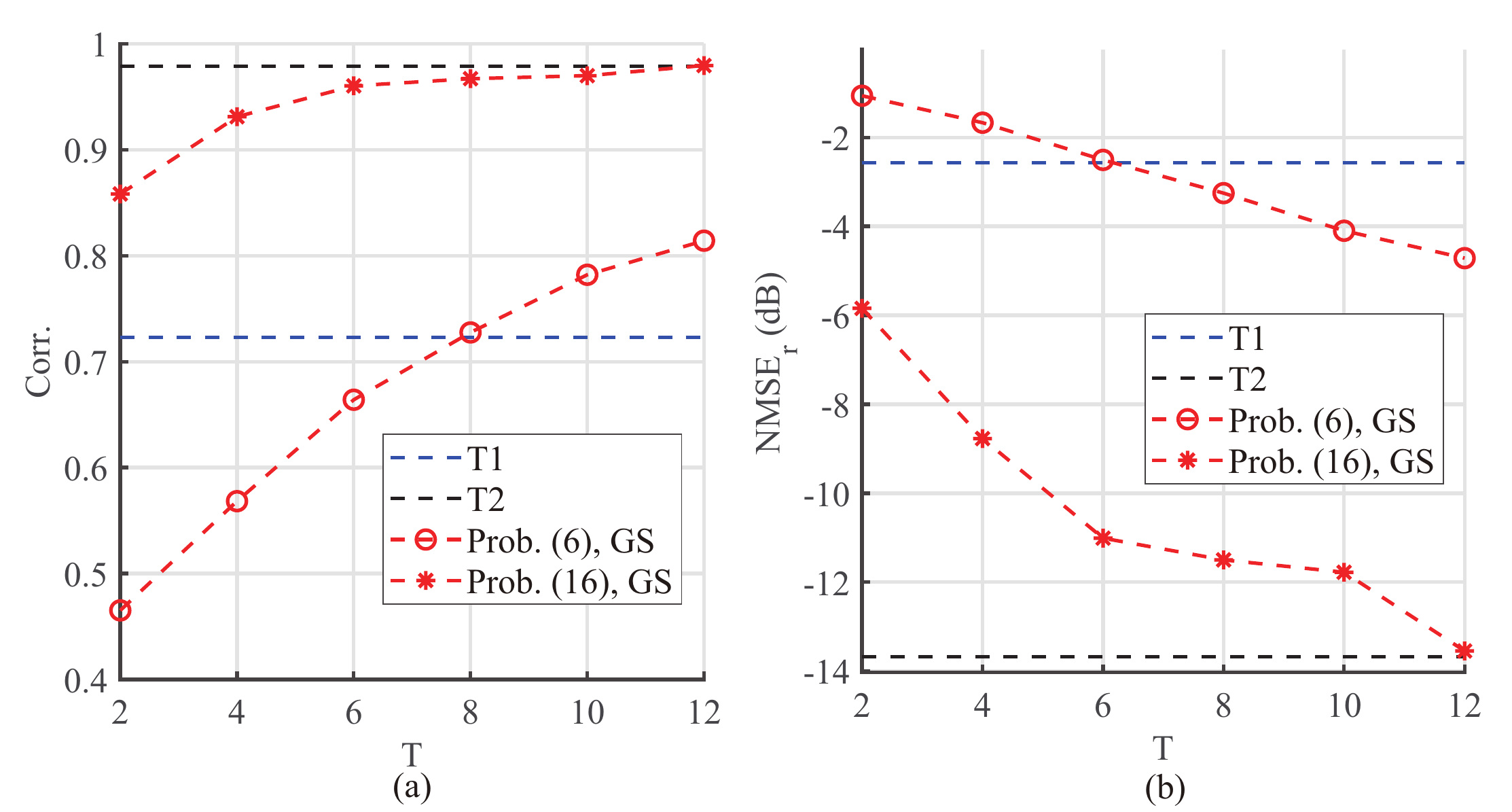}}	
		\caption{The averaged performance comparison of `Prob. (\ref{eq:PR}), GS' and `Prob. (\ref{eq:PR_g}), GS'. `T1' and `T2' stand for the performance of the Type-I and Type-II codewords, respectively.} 
		\label{fig:cmp_g_h}
	\end{figure}
	
	
	Fig. \ref{fig:cmp_g_h} shows the averaged performance of the two schemes. `Prob. (\ref{eq:PR}), GS' stands for the scheme which solves problem (\ref{eq:PR}) with $\hb$, while `Prob. (\ref{eq:PR_g}), GS' stands for the scheme which solves problem (\ref{eq:PR_g}) with $\gb \in \Cbb^5$ firstly and then constructs $\hb^* = \Db \gb^*$. `GS' represents the random precoding sampled from the complex Gaussian distribution. One can see that the performance of `Prob. (\ref{eq:PR_g}), GS' is elevated quickly with slightly increased $T$, while that of `Prob. (\ref{eq:PR}), GS' is improved slowly. The correlation achieved by `Prob. (\ref{eq:PR_g}), GS' is always better than that of the Type-I codeword. However, `Prob. (\ref{eq:PR}), GS' needs about $T = 8$ rounds of feedback to achieve a higher correlation than the Type-I codeword. Even with $T = 12$, there is still a $9~\text{dB}$ gap on $\text{NMSE}_r$ between the two schemes. The results are reasonable. As analyzed in \cite{PR_4N_1, PR_4N_2}, the variable with a larger dimension in problem (\ref{eq:PR}) requires more rounds of feedback to get a good sensing performance. Through our dimension reduction design, the dimension of $\gb$ is much smaller than the orginal $\hb$, alleviating the requirement of feedback. In addition, it verifies our proposed basis matrix construction enables a good sensing recovery $\gb$ to $\hb$.
	
	\subsubsection{\bf Hybrid precoding design} The proposed hybrid precoding scheme is compared with the `GS' precoding scheme. Both the two schemes solve problem (\ref{eq:PR_g}). 
	
	\begin{figure}[t] 
		\centering	
		{\includegraphics[width=0.85\textwidth]{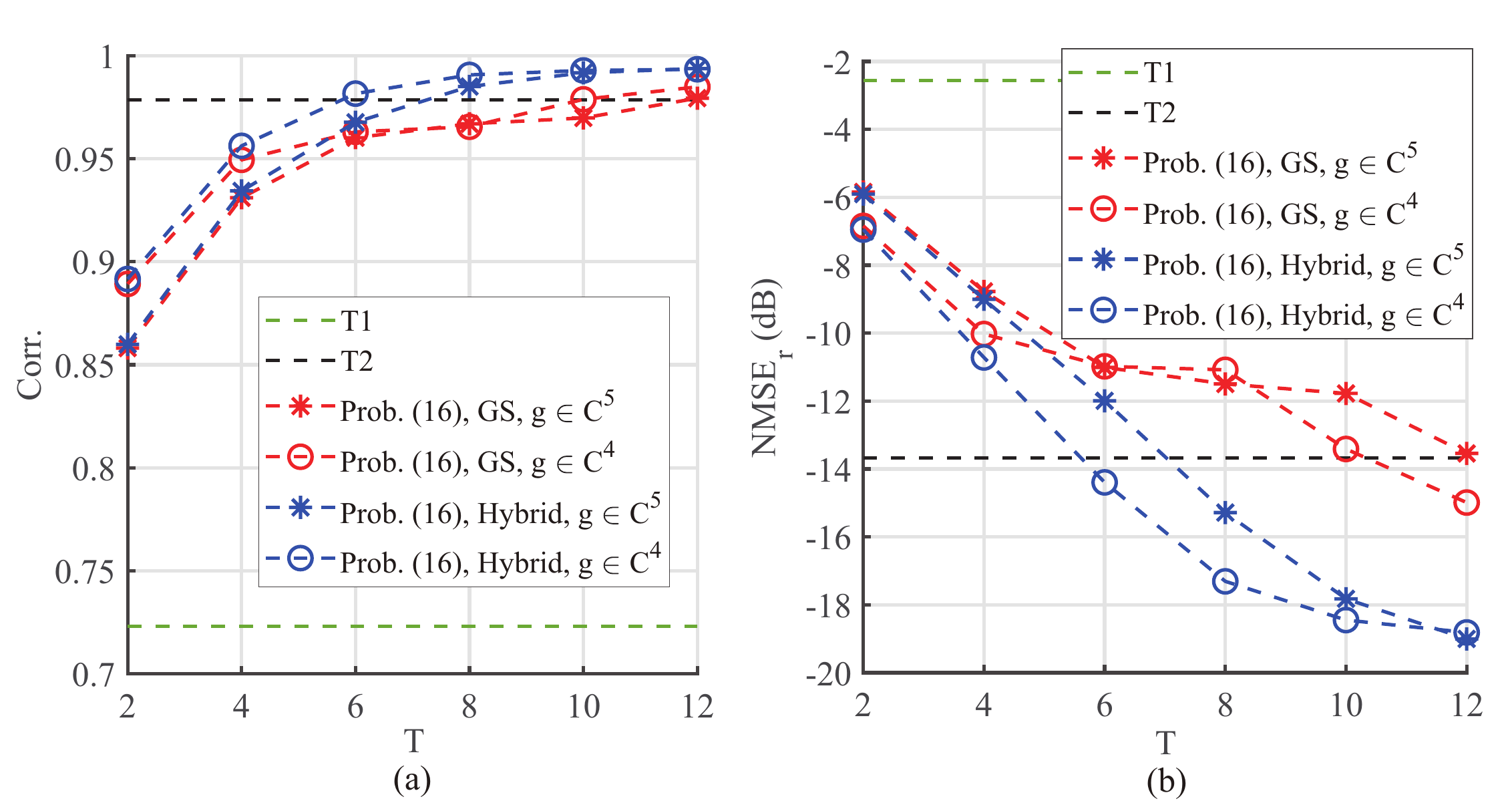}}
		\caption{The averaged performance comparison of the conventional random precoding and that of the proposed hybrid precoding. } 
		\label{fig:cmp_GS_hyb}
	\end{figure}

	Fig. \ref{fig:cmp_GS_hyb} demonstrates that the `hybrid' scheme outperforms the `GS' scheme, under different $L$. The reason behind is that the proposed $\Wb_{t2}$ in the hybrid precoding scheme can concentrate the measurements more on the major components of the CSI. Therefore, it alleviates the requirement of feedback to attain a good sensing performance in the wireless environment with copious propagation paths. 
	
\begin{figure}[htbp] 
		\centering	
		{\includegraphics[width=0.85\textwidth]{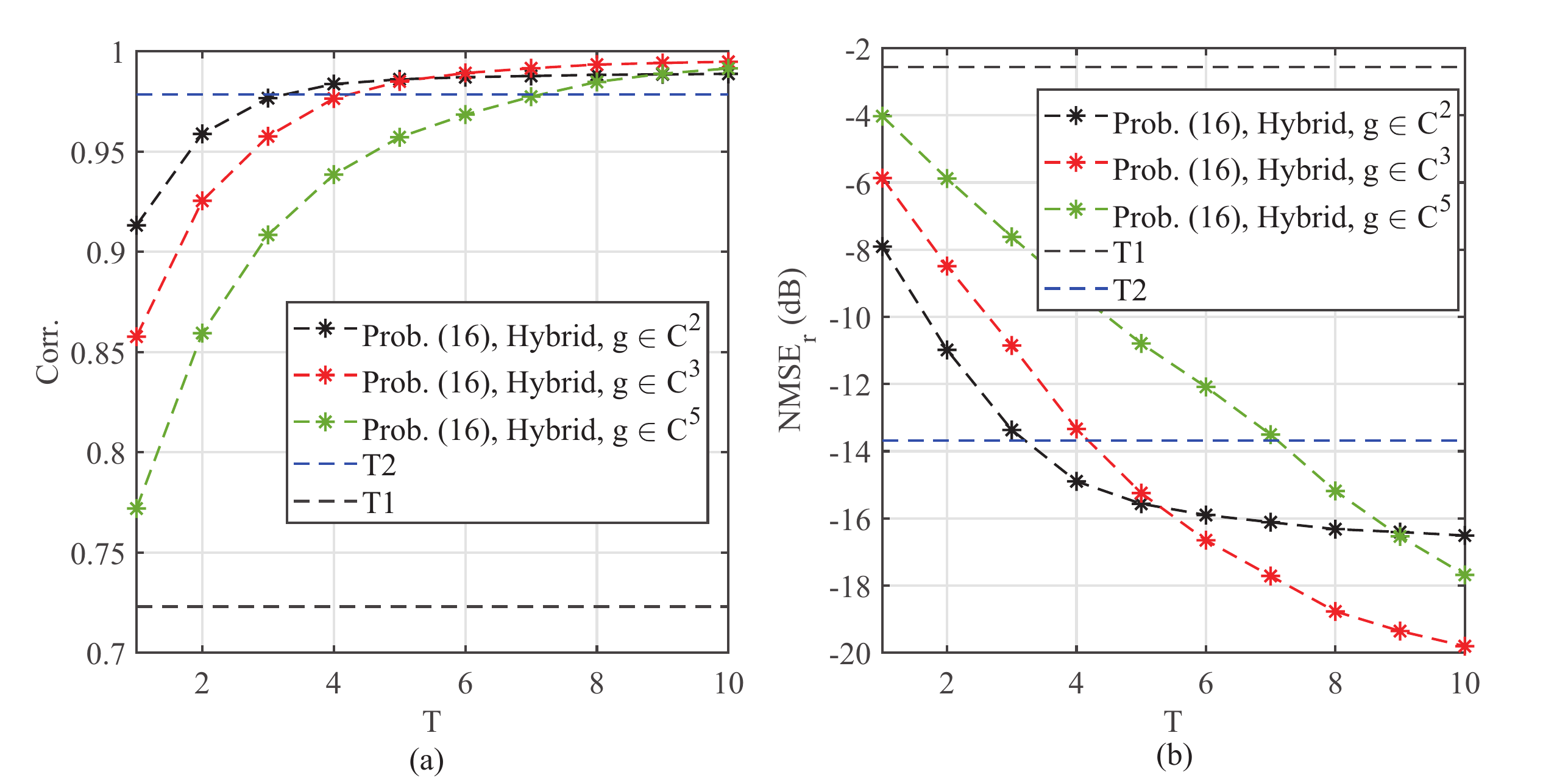}}	
		\caption{The performance of the hybrid precoding scheme under different dimensions of $\gb$.} 
		\label{fig:cmp_g_dim_1}
	\end{figure}

	Besides, the impact of the $\gb$'s dimension $L$ is also tested. Fig. \ref{fig:cmp_g_dim_1} compares the sensing performance of the proposed hybrid precoding scheme under different values of $L$. {It shows that the performance with a smaller value of $L$ is elevated more quickly. The scheme with $\gb \in \Cbb^2$ achieves the best performance for $T\le 5$. This is because it is easier to recover a low-dimensional $\gb$ in (\ref{eq:PR_g}) with high accuracy under limited rounds of feedback. In comparison, the schemes with a larger dimension $L$ achieve the better performance when $T$ is larger, as they can better approximate the channel and characterize it when the number of feedbacks is sufficient to guarantee a good recovery of $\gb$.  }
	
	\subsubsection{\bf CPR formulation} 
\begin{figure}[t] 
		\centering	
		{\includegraphics[width=0.85\textwidth]{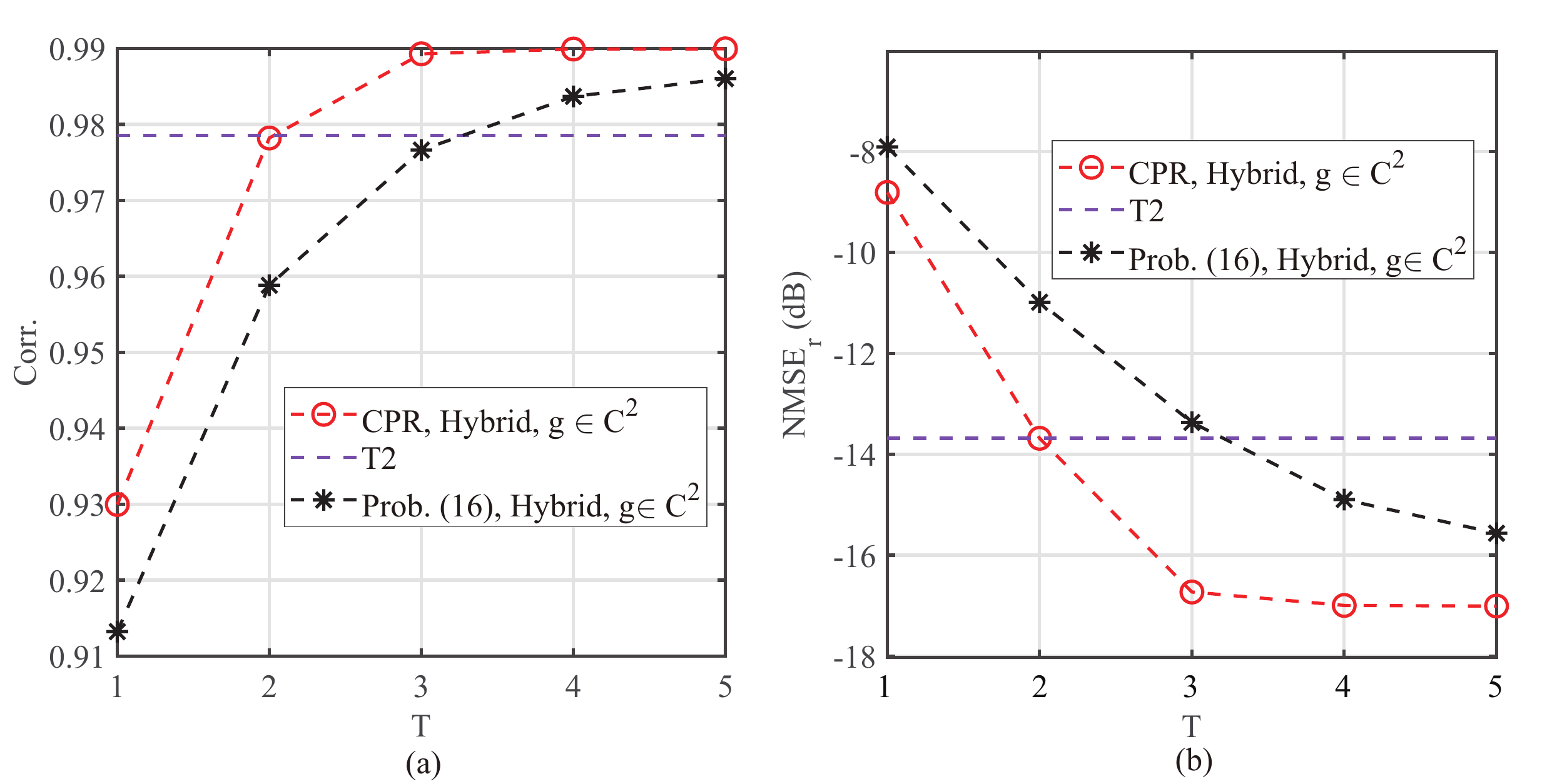}}	
		\caption{The performance of the proposed CPR formulation versus $T$.} 
		\label{fig:cmp_epr}
	\end{figure}

As shown in Fig. \ref{fig:cmp_g_dim_1}, the hybrid scheme with $\gb \in \Cbb^2$ only requires approximately $T = 3$ feedback rounds to get the performance near that of the Type-II codeword. So far, it achieves the best performance for small $T$s. It is further compared in Fig. \ref{fig:cmp_epr} with our proposed CPR formulation (\ref{p:CPR}) solved by the proposed PD-EVD algorithm. One can see that the CPR solved by the PD-EVD algorithm can further boost the sensing performance. It takes only $T= 2$ rounds of feedback to achieve the performance of the Type-II codeword, and reduces the overhead effectively. By shrinking the feasible set of the solution, the CPR scheme enables the very limited feedback rounds ($T = 1, 2$), which is appealing to the resource saving for communication in practice. 
	
	\subsubsection{\bf CQI quantization and algorithm comparison}
	
	\begin{figure}[htb] 
		\centering	
		\includegraphics[width=0.85\textwidth]{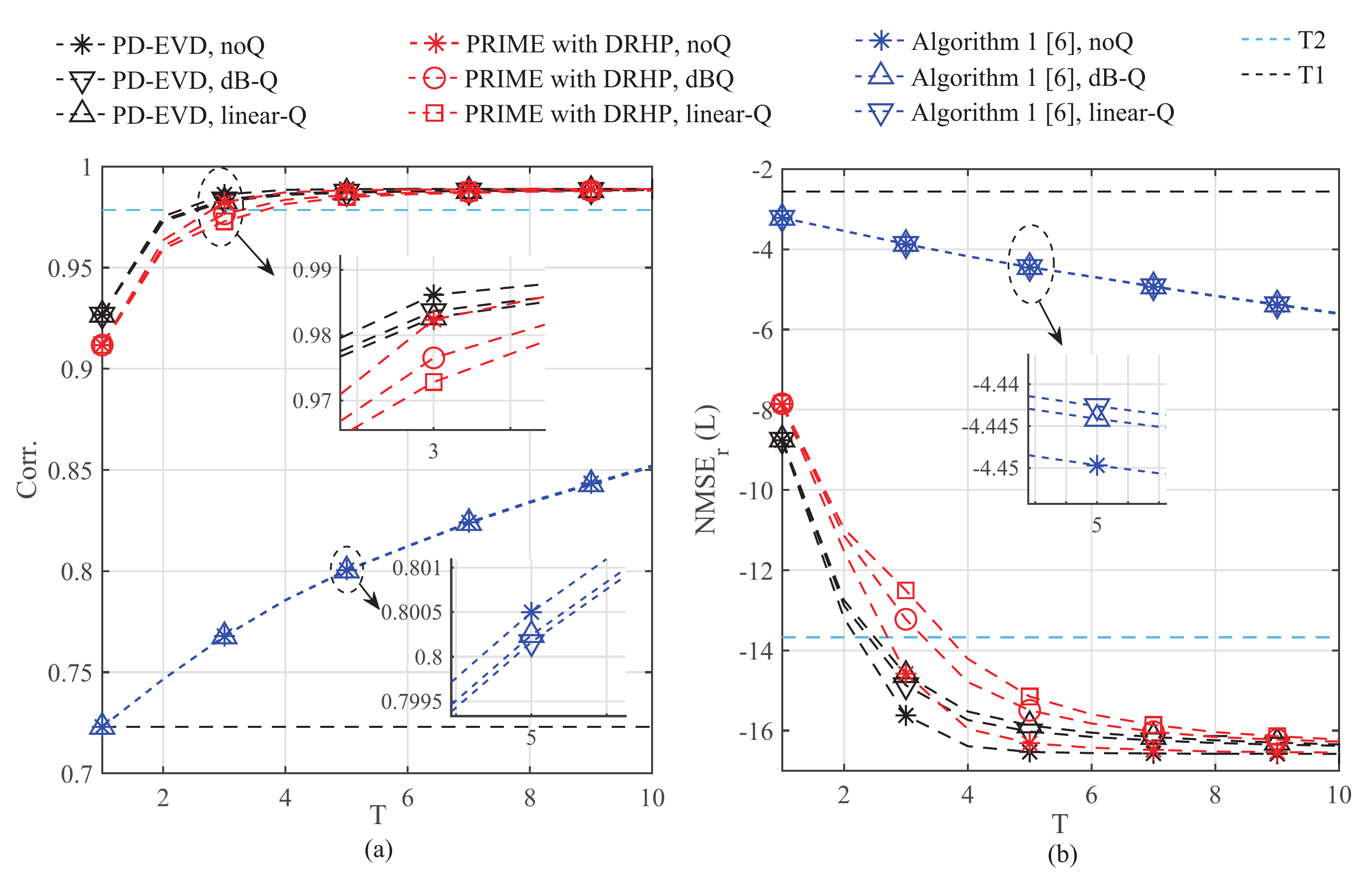}
		\caption{The performance comparison of different algorithms, under different CQI quantization schemes. } 
		\label{fig:cmp_Q} 
	\end{figure}
	
	In practical systems, the original CQI value is quantized before feeding back to the BS. Specified in the 3GPP TS 38.214 \cite{TS_38214}, the CQI is quantized with 4 bits. Here the impact of quantization to our proposed hybrid precoding scheme is tested. The CQI range $[q_{t, min}, q_{t, max}] = [3.35, 28.89]$ is collected from all the $n_{TU} T n_{GS} = 100,000$ CQI values. Note that all original CSI $\hb$ in the test is normalized to $\| \hb \|_2 = 1$. 	
	In practice, the strong reciprocity of the power of the DL-UL CSI usually holds, even for the FDD systems \cite{TWC_JS}. Therefore, the amplitude ambiguity of the reconstructed CSI can be solved by the reciprocity. 
	
	In the experiments, two different quantization schemes are tested. The first one is linear quantization, and the other one is logarithmic quantization denoted by `dB-Q'. Given $n_b$ bits, linear quantization directly quantizes the original CQI values with $2^{n_b}$ uniform intervals. Logarithmic quantization firstly transforms the original CQI values to that with dB unit and conducts uniform quantization in the logarithmic domain. 
	
	Fig. \ref{fig:cmp_Q} shows the performance of CPR solved by our proposed PD-EVD. The performance of PRIME \cite{PRIME_tsp16} and Algorithm 1 proposed in \cite{LiKai_2} is also shown for comparison. Note that Algorithm 1 in \cite{LiKai_2} solves problem (\ref{eq:PR}), while PRIME solves problem (\ref{eq:PR_g}) with our proposed dimension reduction and hybrid precoding (DRHP) applied. For CQI quantization, it shows that the performance of Algorithm 1 \cite{LiKai_2} with quantization and that without quantization are almost the same, while a little performance loss is caused to PD-EVD and PRIME. Comparing `dB-Q' with `linear-Q', the former one has the better performance. For the sensing performance of the three algorithms, PD-EVD outperforms the other two algorithms. It only takes $T=3$ to exceed the performance of Type-II codewords, no matter the CQI values are quantized or not. In comparison, the performance of Algorithm 1 \cite{LiKai_2} is only slightly better than that of the Type-I codewords. It is necessary to point out that the performance of PRIME without DRHP applied will degrade a lot and almost coincide with that of Algorithm 1 \cite{LiKai_2}. Moreover, it is found in our tests that Algorithm 1 \cite{LiKai_2} and PRIME without DRHP requires $T > 20$ rounds of feedback to get the comparable performance as Type-II codewords. 
	
	
	\subsection{Performance on QuaDriGa Dataset}
	In the section, the proposed CSI sensing schemes are tested over the CSI generated by QuaDriGa. The BS is located at $(x, y) = (0, 0)~(m)$. Totally 2000 UE samples are randomly located in the $30 \times 30~(m)$ square area centered at $(x, y) = (215, 0)~(m)$. Here a slightly different approach is applied to model RU. That is, given each TU from the whole samples, the other UE samples are treated as the RU dataset. If otherwise specified, $n_R = 10$ RUs are selected from the nearest one to the furthest one to the TU. Different from that in DeepMIMO wherein the propagation environment is mainly line-of-sight (LoS), the non-line-of-sight (NLoS) scenario is considered in QuaDriGa. The number of propagation paths of the CSI generated in the NLoS scenario is $n_L = 61$. The parameters to generate the CSI are listed in Table \ref{Tb:simu_para}.
	
\begin{figure}[htbp]
		\centering	
		{\includegraphics[width=0.85\textwidth]{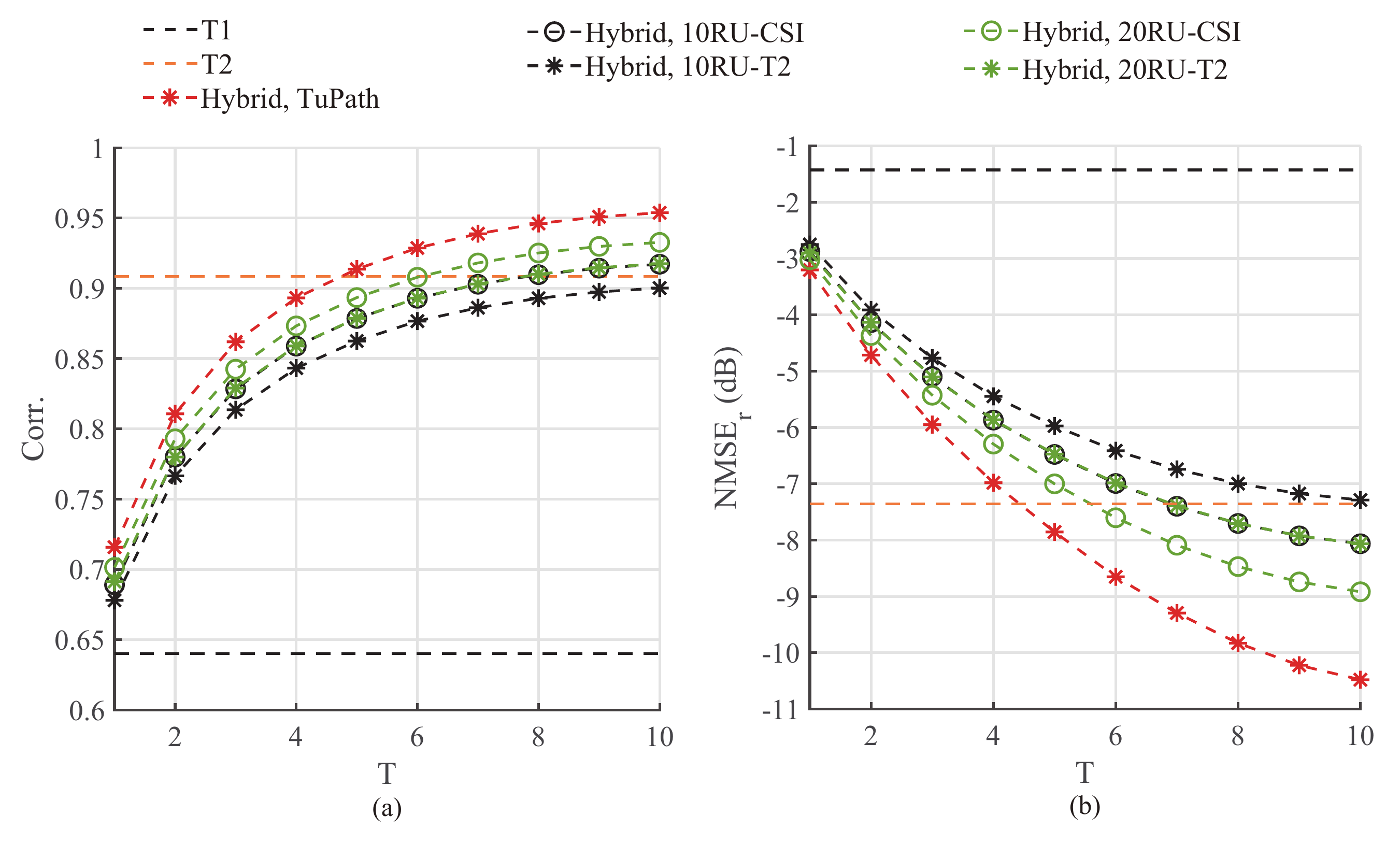}}
		\caption{The performance comparison of different schemes to construct $\Db$, in the NLoS scenario.}
		\label{fig:cmp_prime_NLoS}
	\end{figure}

	Three different approaches to construct the basis matrix $\Db$ in (\ref{eq:D_pca}) are considered and the performance is shown in Fig. \ref{fig:cmp_prime_NLoS}. `RU-CSI' conducts the SVD in (\ref{eq:C_h}) based on the CSI in (\ref{eq:H}), while `RU-T2' is based on the Type-II codewords as (\ref{eq:Cov_T2}). In `TuPath', the path coefficients of the TU are used to compute the SVD in (\ref{eq:C_h}). PRIME is applied in all the three schemes to solve problem (\ref{eq:PR_g}) with $\gb \in \Cbb^5$.
	
	
	\begin{table*}[t]
		\centering
		\caption{The parameters to generate the CSI in QuaDriGa.} \label{Tb:simu_para}
		\begin{tabular}{l|llllll}   \hline
			Parameter & Value \\ \hline
			Propagation scenario & 3GPP 38.901 UMa (Urban Macro-Cell)\\
			BS height &  25 meter\\
			UE height & 1.5 meter \\
			Antenna model at UE & Omni \\
			Number of antennas at the BS, $M = 2 N_{h} N_{v}$ & 32 \\
			Number of antennas in the horizontal dimension, $N_{h}$ & 8 \\ 
			Number of antennas in the vertical dimension, $N_{v}$ & 2 \\
			Antenna horizontal spcaing & $0.5 \lambda$ \\
			Antenna vertical spcaing & $ 4 \times 0.7 \lambda$ \\
			Elevation tilt angle &  $7^{\circ}$ \\
			Center carrier frequency $f$ (GHz) & 2.16\\ 
			\hline
		\end{tabular}
				\vspace{0.3cm}
	\end{table*}
	
	%
	%
	
	Fig. \ref{fig:cmp_prime_NLoS} shows that `TuPath’ attains the best performance. From ‘RU-CSI’ to ‘RU-T2’, the performance gets worse, since the Type-II codeword only conveys approximate information of the original CSI of RU. It also demonstrates that the performance of `RU-T2' is still lower than the Type-II codeword with $T= 10$ in the NLoS scenario. Comparing `TuPath' with `RU-T2', it can be inferred that the performance loss is mainly due to the inaccuracy of the basis matrix. To handle it, the number of RUs to construct the basis matrix is increased to 20. It shows in Fig. \ref{fig:cmp_prime_NLoS} that the performance of `20RU-T2' is improved and exceeds the Type-II codeword for $T\ge 8$.
	
	\begin{figure}[htbp]
		\centering	
		{\includegraphics[width=0.85\textwidth]{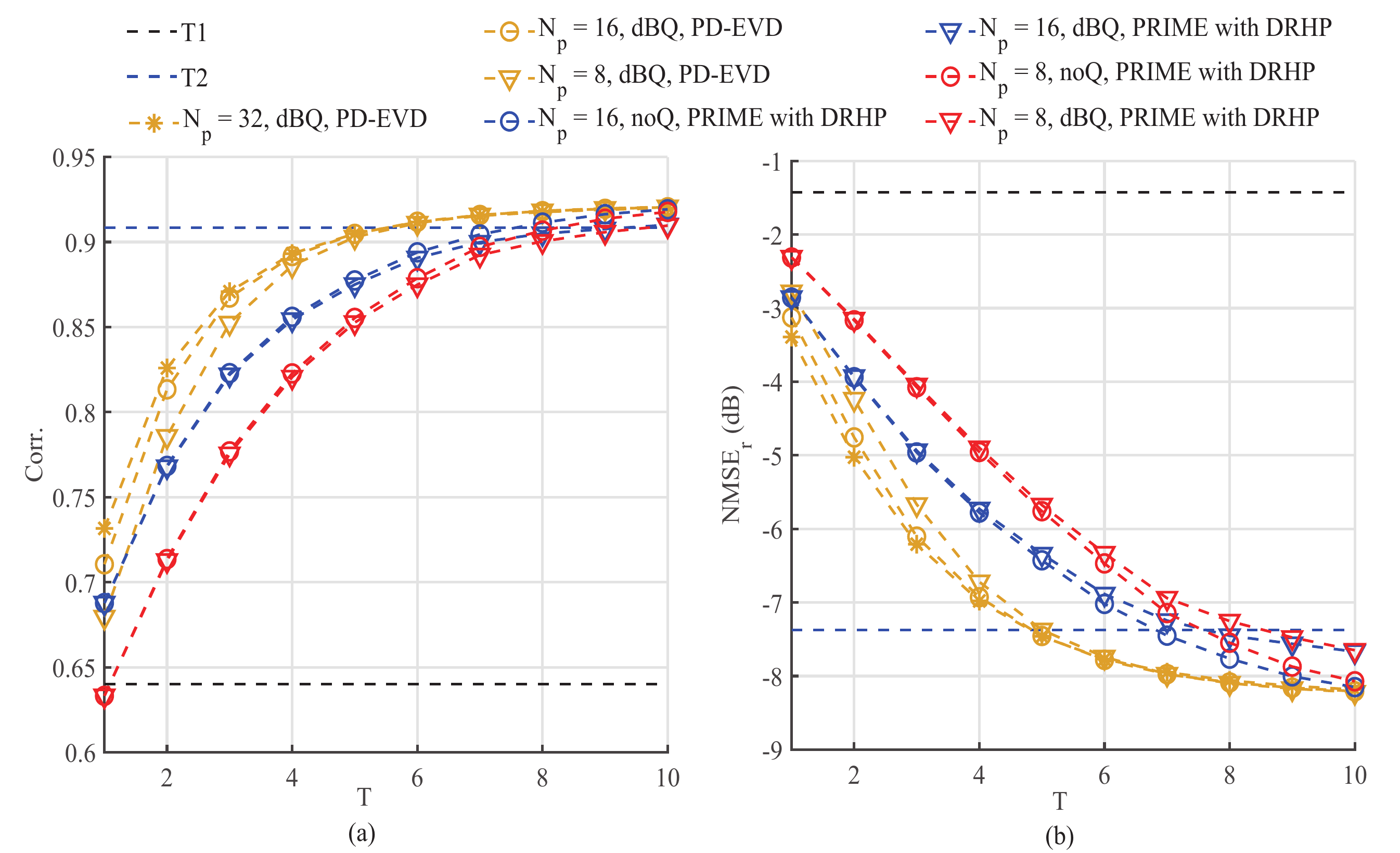}}
		\caption{The performance comparison of CPR by PD-EVD and that of PRIME with DRHP, under different values of $N_p$.}
		\label{fig:cmp_diff_csi_port_EPR}
	\end{figure}
	
	
	Fig. \ref{fig:cmp_diff_csi_port_EPR} compares the performance of our proposed PD-EVD to solve CPR and that of PRIME to solve problem (\ref{eq:PR_g}). Fewer CSI ports $N_p$ are taken into account and 4-bit `dB-Q' CQI quantization is applied.  One can observe that our proposed CPR with PD-EVD lifts the performance effectively. Whatever the number of CSI ports $N_p$ is applied, the CPR with PD-EVD only takes about $T = 5$ rounds of feedback to reach the performance of the Type-II codeword. 
	\FloatBarrier
	\subsection{MECS-SGDA versus PD-EVD}

	\begin{figure}[htbp]
		\centering	
		\includegraphics[width=.85\textwidth]{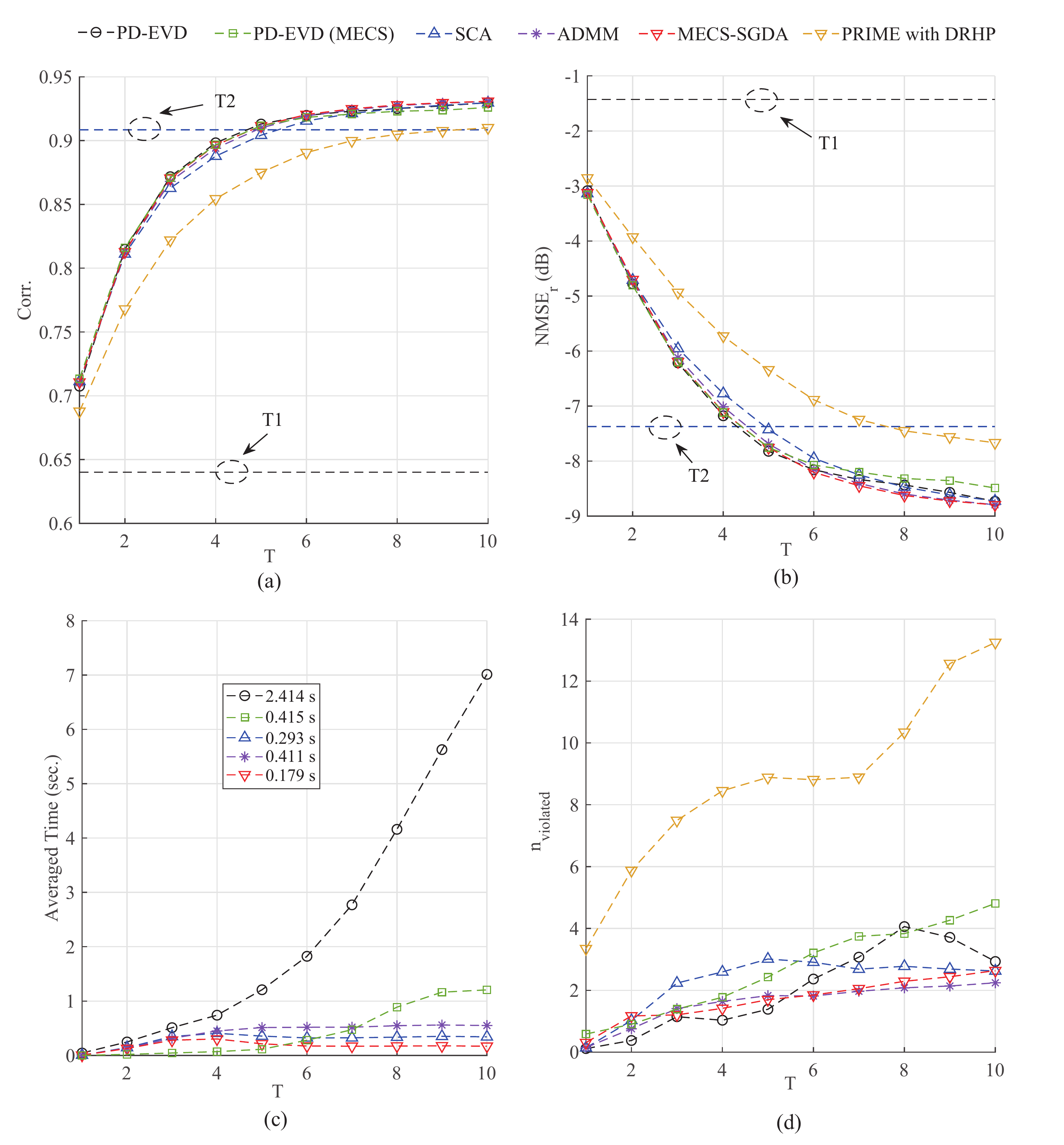}
		\caption{The performance comparison on CSI sensing, computational time and number of violated constraints among MECS-SGDA, PD-EVD, ADMM and SCA. CQIs are quantized by `4bits-dBQ' and $N_p = 16$. The labels in subfig (c) mark the averaged computational time. `PRIME with DRHP' that solves the unconstrained problem (\ref{eq:PR_g}) is adopted as a benchmark. }
		\label{fig:cmp_all_4bit_dBQ}
		\vspace{0.3cm}
	\end{figure}

	In the section, the CSI sensing performance and the computational time are evaluated for PD-EVD and the two-stage MECS-SGDA algorithm (Algorithm \ref{alg:MECSC} + Algorithm \ref{alg:GDA}). For comparison, the performance of a successive convex approximation (SCA) algorithm based on CCP \cite{park2017general} and that of an consensus-ADMM methed modified from \cite{HKJ} to solve the CPR problem are also presented. SGDA, SCA and ADMM solve problem (\ref{p:orig}) with the MECS obtained by our proposed Algorithm \ref{alg:MECSC}. PD-EVD solves problem (\ref{p:CPR_g_SDR}) over the overall constraint set, while PD-EVD (MECS) solves problem (\ref{p:CPR_g_SDR}) over the obtained MECS.  The experiments are conducted with MATLAB 2021 on a computer with Intel Core i9-10900X CPU and 32 GB RAM.
	
	Fig. \ref{fig:cmp_all_4bit_dBQ} (a) and (b) show the CSI sensing performance of these four algorithms. One can observe that the obtained correlation and $\text{NMSE}_r$ of SGDA, PD-EVD, SCA and ADMM are almost the same. For the averaged computational time shown in  Fig. \ref{fig:cmp_all_4bit_dBQ} (c), one can see that PD-EVD (MECS) takes much less time (0.415 s) than PD-EVD (2.414 s). It is in accordance with the expectation, since the number of constraints has been greatly reduced from the original constraint set through the proposed MECS construction. Moreover, among all the algorithms with MECS, it shows our proposed two-stage MECS-SGDA is the most computationally efficient one (0.179 s). The reason is that the proposed SGDA Algorithm \ref{alg:GDA} only involves one-loop iterations rather than double-loop iterations as in SCA, and its update in each iteration does not take complicated computations as in ADMM. Combining the results from Fig. \ref{fig:cmp_all_4bit_dBQ} (a) to (c), one can see that our proposed  MECS-SGDA scheme achieves the best performance on both the CSI accuracy and the computational time. 
	
	{In Fig. \ref{fig:cmp_all_4bit_dBQ} (d), the averaged number of unsatisfied constraints of the solutions returned by these algorithms is plotted. It shows that the returned solutions by these algorithms satisfy most of the constraints. Among these hundreds to thousands of constraints, only several ones are not met. While the number of violated constraints by the solution of `PRIME with DRHP' for the unconstrained PR (\ref{eq:PR_g}) seems not very large, it is found in experiments that its averaged violation defined as $\sum_{j, t} \max \big(|\ub_{j}^H  \Wb_t^H \hb^*|^2 - | \ub_{j_t^*}^H  \Wb_t^H \hb^*|^2, 0 \big)$ is 5.03, which is more than one-order larger than that of SGDA (0.41). Therefore, `PRIME with DRHP' suffers from a noticeable performance loss in terms of both correlation and $\text{NMSE}_r$, as shown in Fig. \ref{fig:cmp_all_4bit_dBQ} (a) and (b). }
	
	
	\FloatBarrier
	\subsection{Performance of Multi-Carrier Systems}
	\begin{figure}[htbp]
		\centering	
		{\includegraphics[width=0.85\textwidth]{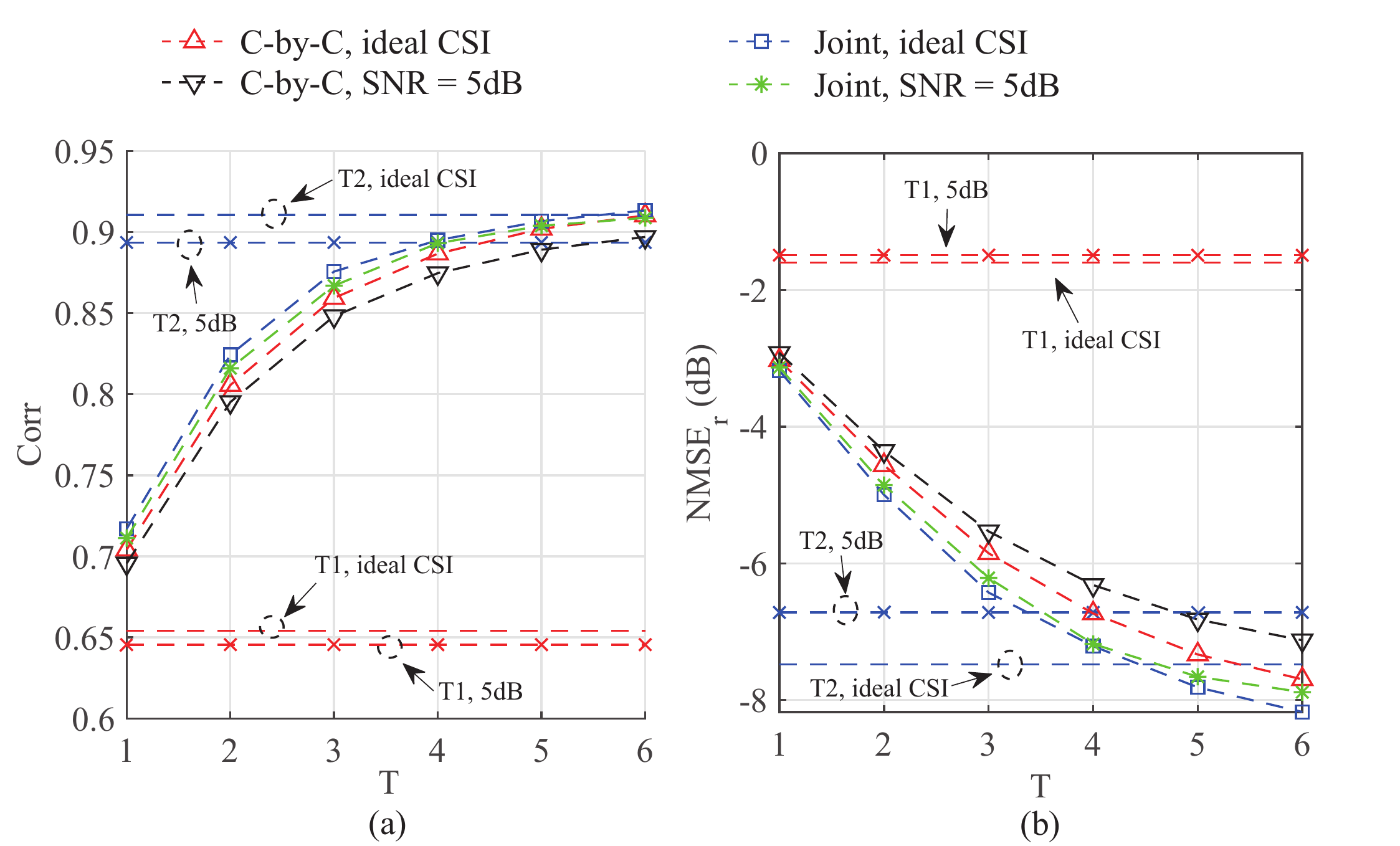}}
		\caption{The performance of carrier-by-carrier sensing and joint sensing.}
		\label{fig:mulC_CbyC}
	\end{figure}

	\begin{figure}[t]
		\centering	
		{\includegraphics[width=0.85\textwidth]{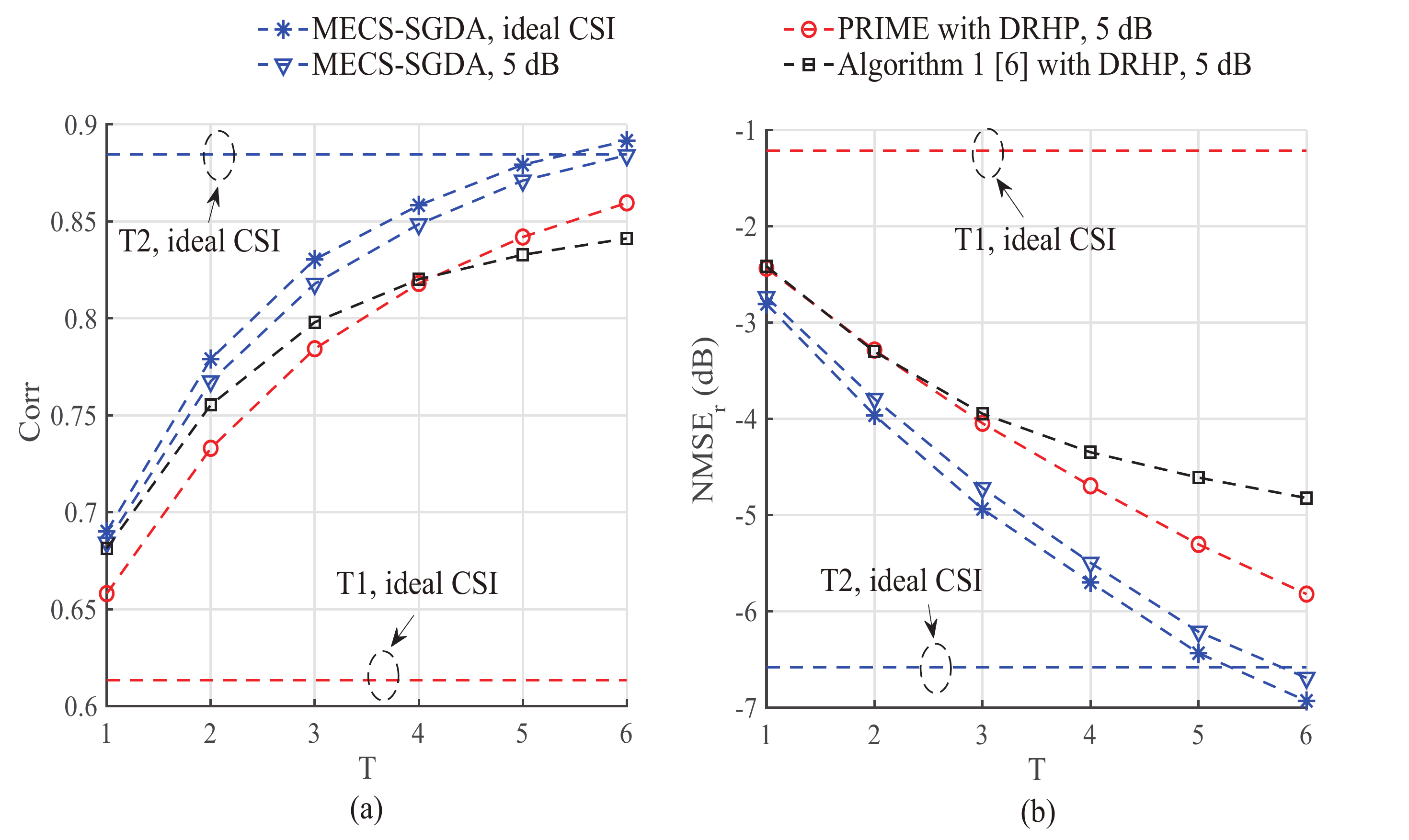}}
		\caption{The performance comparison on SGDA, PRIME and Algorithm 1 \cite{LiKai_2} for multi-carrier sensing.}
		\label{fig:mulC_SNR}
	\end{figure}
	
	
	In this section, the performance of our proposed  MECS-SGDA is tested for the multi-carrier scenario. The performance of the carrier-by-carrier (C-by-C) sensing and that of the joint sensing in Section \ref{sec:mulC} are compared firstly. In the former one, the CSI over each carrier is recovered one-by-one via solving problem (\ref{p:CPR}), while in the latter one the CSI over all carriers is recovered jointly by solving problem (\ref{p:CPR_mulC}). The frequency band of each UE is consisted of $n_C = 12$ sub-carriers. The value $L$ in the C-by-C sensing and the values of $\tilde{L}, \tilde{n}_L$ in the joint sensing are all set to be 5. The PMI and CQI are fed back per carrier. The number of CSI ports is $N_p = 16$ and the size of the Type-I codebook is $n_{T1} = 512 = 2^9$. `Idea CSI' means that the effective CSI to select the PMI and compute the CQI is ideal without considering the DL channel estimation error  in (\ref{eq:CSI_eff}), while `SNR = 5dB' means that the CSI is estimated via the LMMSE estimation under 5dB SNR setting. The CQI is quantized with `4bits-dBQ'.
	
	Fig. \ref{fig:mulC_CbyC} demonstrates the performance of `C-by-C' sensing and that of the joint sensing. One can observe that the joint sensing outperforms the `C-by-C' sensing, no matter the DL CSI is ideal or with channel estimation error. Moreover, the performance of the joint sensing with channel estimation error is very close to that with ideal CSI. Even with the CSI estimation error, the joint sensing can exceed the performance of Type-II codewords with $T=4$ rounds of feedback, while the `C-by-C' sensing requires $T = 6$ rounds. The performance improvement is reasonable, as the frequency correlation of the CSI over different carriers is further utilized besides the angular correlation in the joint sensing.
	
	In Fig. \ref{fig:mulC_SNR} the joint sensing is tested under a more flexible feedback setting. The number of carriers of each UE is $n_C = 48$. Different feedback granularity over sub-carriers are chosen for PMI and CQI. One PMI is fed back over every $12$ sub-carriers, while one CQI with `4bit-dBQ' is returned over every $4$ sub-carriers. The SNR in the DL channel estimation (by LMMSE) to obtain the effective CSI is $5~\text{dB}$. The performance of PRIME and Algorithm 1 \cite{LiKai_2} with DRHP is also shown for comparison. One can observe that our proposed MECS-SGDA achieves the better CSI sensing performance than PRIME and Algorithm 1 \cite{LiKai_2}. With the DL CSI estimated SNR =  $5~\text{dB}$, there is a slight performance loss. It is also found in the experiments that the performance gap is almost closed when the SNR is increased to $10~\text{dB}$. Moreover, our proposed scheme takes $T = 6$ rounds of feedback to acquire the better performance than that of the Type-II codeword. While the feedback overhead of the Type-II codeword is $20 + 12\times 56 = 692$ bits, our proposed scheme takes $(4\times9 +12 \times 4)T = 504$ bits, bringing approximately $27\%$ overhead saving. 
	
	\FloatBarrier
	\section{Conclusion \label{sec:conclusion}}
	
	In this paper, we have investigated the CSI sensing in heterogeneous networks consisting of UEs with different feedback capabilities. 
	We have proposed a new DL CSI sensing scheme with only slight feedback overhead.  
	Our design includes: (i) a novel parameter dimension reduction method by exploiting the spatial consistency of nearby UEs of the TU; (ii) a hybrid precoding design to balance the sensing diversity and the SNR enhancement at the UE, (iii) an CPR formulation which effectively characterizes the solution set of the CSI via the PMI feedback,  and (iv) a fast two-stage algorithm that combines the MECS construction and the first-order SGDA algorithm. 
	Moreover, the proposed CSI sensing scheme has been extended to the multi-carrier systems which can yield improved performance thanks to the correlation of the CSI across different subcarriers.
	Through extensive experiments based on practical datasets, we have shown that our proposed design can effectively elevate the CSI sensing performance of the TU, close the performance gap between the Type-I and the Type-II codewords and bring significant feedback overhead reduction.
	
	\footnotesize 
	\bibliographystyle{IEEEtran}
	
	\bibliography{refs_journal}

\end{document}